\documentclass[12pt]{article}
\usepackage{amsfonts,amsmath}
\usepackage{amssymb}

\makeatletter \@addtoreset{equation}{section} \makeatother

\def\theequation{\thesection.\arabic{equation}}

\newcommand{\XX}{{X^\prime\,{}}}
\newcommand{\XXX}{{X^{\prime\prime}\,{}}}
\newcommand{\gY}{{\mathcal{Y}}}

\newcommand{\Z}{{{\cal Z}}}

\newcommand{\W}{\mathcal{W}}
\newcommand{\U}{\mathcal{U}}

\newcommand{\dete}{s}
\newcommand{\cdete}{\bar{s}}

\newcommand{\cZ}{{\overline{\mathcal{Z}\rule{0pt}{10pt}}}}
\newcommand{\cY}{{\overline{\gY\rule{0pt}{10pt}}}}

\newcommand{\bw}{{{w}}}

\newcommand{\Ds}{\mathfrak{S}}

\newcommand{\Hh}{{\cal H}}
\newcommand{\B}{{\cal B}}

\newcommand{\tudasuda}{{\circ}}
\newcommand{\vga}{A}
\newcommand{\vgb}{B}
\newcommand{\vy}{Y}

\newcommand{\ie}{{{\it i.e.,} \ }}

\newcommand{\ZIG}{\mathfrak{H}}
\newcommand{\ZIGM}{\mathfrak{H}_M}
\newcommand{\GZM}{\mathfrak{H}_M\times\mathbb{C}^M}

\newcommand{\be}{\begin{equation} }

\newcommand{\ee}{\end{equation}}

\newcommand{\bee}{\begin{eqnarray}}

\newcommand{\beee}{\begin{array}}
\newcommand{\eee}{\end{eqnarray}}
\newcommand{\eeee}{\end{array}}

\newcommand{\gn}{\nu}

\newcommand{\gm}{\mu}

\newcommand{\gx}{\xi}
\newcommand{\gsing}{\Lambda}
\newcommand{\gr}{\rho}
\newcommand{\ga}{\alpha}
\newcommand{\gb}{\beta}
\newcommand{\gga}{\gamma}

\newcommand{\M}{{\cal M}}
\newcommand{\Mi}{{ M^d}}
\newcommand{\T}{{\cal T}}
\newcommand{\A}{{\cal A}}

\newcommand{\ls}{\!\!\!\!\!\!}

\newcommand{\gd}{\delta}
\newcommand{\gl}{\lambda}
\newcommand{\gk}{p}
\newcommand{\gep}{\epsilon}
\newcommand{\gvep}{\varepsilon}

\newcommand{\gs}{\sigma}

 \newcommand{\bu}{u} 

\newcommand{\q}{\,,\qquad}

\newcommand{\done}{{{1}^\prime}}
\newcommand{\dtwo}{{{2}^\prime}}

\newcommand{\dm}{{{\gm^\prime}}}

\newcommand{\da}{{{\ga^\prime}}}
\newcommand{\db}{{{\gb^\prime}}}
\newcommand{\dd}{{{\gd^\prime}}}
\newcommand{\dga}{{{\gga^\prime}}}

\newcommand{\X}{{  \mathbf{X}}}
\newcommand{\Y}{{\mathbf{Y}}}
 
  \newcommand{\go}{\omega}

\newcommand{\bZ}{{\overline{Z}}}

\newcommand{\nn}{\nonumber}
\newcommand{\half}{\frac{1}{2}}

\newcommand{\p}{\partial}

\newcommand{\D}{{\cal D}}
\newcommand{\R}{{\cal R}}

\newcommand{\C}{{\cal C}}

\renewcommand{\S}{{\cal S}}
\newcommand{\f}{\frac}
 \newcommand{\dis}{\displaystyle}

\newcommand{\za}{{{A}}}
 \newcommand{\zb}{{{B}}}
\newcommand{\zga}{{{A}}}

\begin{document}

\begin{flushright}
\vspace{1mm} FIAN/TD/12-08\\
January {2008}\\
\end{flushright}

\begin{center}
{\large\bf
 Higher Spin Fields in Siegel Space, Currents and\\ Theta Functions}
\vglue 0.6  true cm

O.A. Gelfond$^1$ and M.A.~Vasiliev$^2$ \vglue 0.3  true cm

${}^1$Institute of System Research of Russian Academy of Sciences,\\
Nakhimovsky prospect 36-1, 117218, Moscow, Russia

\vglue 0.3  true cm

${}^2$I.E.Tamm Department of Theoretical Physics, Lebedev Physical
Institute,\\
Leninsky prospect 53, 119991, Moscow, Russia

\end{center}

\begin{abstract}

Dynamics of four-dimensional massless fields of all spins is
formulated in the Siegel space   of complex $4\times 4$ symmetric
matrices. It is shown that the unfolded equations of free massless
fields, that have a form of multidimensional Schrodinger
equations, naturally distinguish between positive- and
negative-frequency solutions of relativistic field equations,
i.e., particles and antiparticles. Multidimensional Riemann theta
functions are shown to solve massless field equations in the
Siegel space. We establish the correspondence between conserved
higher-spin currents in four-dimensional Minkowski space and those
in the ten-dimensional matrix space. It is shown that global
symmetry parameters of the current in the matrix space should be
singular to reproduce a nonzero current in Minkowski space. The
$\D-$function integral evolution formulae for 4d massless fields
in the Fock-Siegel space are obtained. The generalization of the
proposed scheme to higher dimensions and systems of higher ranks
is considered.

\end{abstract}

\newpage
\tableofcontents
\newpage
\section{Introduction}\label{Generalities}

The idea that a set of massless fields of all spins in the
four-dimensional space should admit a natural description in the
ten-dimensional space $\M_4$ with real symmetric matrix
coordinates $X^{AB}=X^{BA}$ ($A,B, \ldots = 1,\ldots, 4$)    was
originally proposed by Fronsdal in \cite{F}. Later, the same
conclusion was independently reached in \cite{BLS}. The dynamical
equations in $\M_M$, that for $M=4$ are equivalent to the field
equations for massless fields of all spins in the four-dimensional
Minkowski space $M^4$, are very simple \cite{BHS}. All integer
spin fields of $M^4$ are described in $\M_4$  by a single scalar
field $b(X)$ that satisfies the Klein-Gordon-like equation \be
\label{oscal} \Big ( \f{\p^2}{\p X^{ A B} \p X^{ C D}} -
\f{\p^2}{\p X^{ A C} \p X^{ B D}}\Big ) b(X)=0\,. \ee All
half-integer spin fields are described by a single fermion field
$f_ B(X)$ that satisfies the Dirac-like equation \be \label{ofer}
\f{\p}{\p X^{ A B}} f_ C(X) - \f{\p}{\p X^{ A C}} f_ B(X)
=0\,. \ee The equations (\ref{oscal}) and (\ref{ofer}) were
derived in \cite{BHS} from the system of equations \be
\label{dydy} \left ( \f{\p}{\p X^{ A B}} + \mu \f{\p^2}{\p Y^ A \p
Y^ B}\right ) C(Y|X) =0\,, \ee where $Y^ A$ were treated as
auxiliary commuting variables (the parameter $\mu\neq 0$ is
introduced  for the future convenience). Although the equations
(\ref{oscal})-(\ref{dydy}) were originally considered for $M=4$,
they make sense  for any $M$.

The equations (\ref{dydy}) express the first derivatives with
respect to space-time variables $X^{AB}$ in terms of the fields
themselves. As such, they belong to the class of  {\it unfolded}
partial differential equations (PDE) that, more generally, express
the exterior differential of a set of differential forms in terms
of exterior products of the differential forms themselves. Such a
first-order form of dynamical field equations can always be
achieved by introducing a (may be infinite) set of auxiliary
fields which parameterize all combinations of derivatives of the
dynamical fields that remain non-zero on the field equations. For
example, in the system (\ref{dydy}), the {\it dynamical fields}
are \be b(X)=C(0|X) \ee and \be f_A(X)= \f{\p}{\p Y^A} C(Y|X)\Big
|_{Y=0}\,. \ee As a consequence of (\ref{dydy}), they satisfy,
respectively, the equations (\ref{oscal}) and (\ref{ofer}). All
the fields \be C_{A_1\ldots A_n} (X) = \f{\p^n}{\p Y^{A_1}\ldots
\p Y^{A_n}} C(Y|X)\Big |_{Y=0} \q n>1 \ee are auxiliary, being
expressed via higher $X$--derivatives of the dynamical fields by
virtue of the equations (\ref{dydy}). In  \cite{BHS} it was shown
that the equations (\ref{oscal}) and (\ref{ofer}) along with
constraints that express the auxiliary fields via $X$--derivatives
of the dynamical fields exhaust the content of the unfolded system
(\ref{dydy}). That the equations (\ref{dydy})  formulated in the
ten-dimensional space-time, still describe massless fields in four
dimensions was also shown in \cite{BHS} using the unfolded
dynamics approach (see also Section \ref{unfold}).

Theories in $\M_M$ have been studied in a number of papers from
different perspectives
\cite{Mar,cur,IB,DV,PST,s3,tens2,ZU,BPST,BBAST,EL,FE,FEL,west,EI,33}.
In this paper, we will further study  the higher-spin (HS) theory
in the matrix space.

The main practical goal is to show how HS conserved currents in
$\M_4$ found in \cite{cur} reproduce usual HS conserved currents
in Minkowski space found in \cite{GSV}. The analysis is not
completely trivial since the conserved charges in $\M_4$ contain
an additional integration over one spinning variable. The apparent
difficulty is that the compact spin space is contractible to zero,
hence  implying that the charge must vanish for regular solutions
of the field equations in $\M_4$. A standard way out would be to
integrate over a noncontractible cycle in $\M_M$. (In fact, in its
$ {Sp}(2M)$ invariant compactification which is Lagrangian
Grassmannian \cite{F,Mar}.) We have not been able to proceed along
these lines, however. Instead we will show in this paper that the
Minkowski charge is correctly reproduced by virtue of introducing
a singularity  that effectively makes the integration cycle over
the spinning variable noncontractible. The obtained results may
have  several applications.

First of all, conserved currents determine the lowest order
Noether interactions with the HS gauge fields associated with the
HS symmetries. It is straightforward to introduce cubic
interactions of HS gauge potentials with conserved currents via
replacing the global symmetry parameters $\eta$ in the charge by
the corresponding HS one-form gauge connections. The results of
this paper show that, to reproduce correctly the HS interactions
in the four-dimensional setup, HS potentials in $\M_4$ should
develop a singularity in the spinning directions. In other words,
the results of this paper indicate that there are nontrivial
fluxes in the spinning directions in $\M_4$, that support charges
in Minkowski space.

Another application is that the obtained formula for conserved
charges allows us to write the integral  evolution representation
for solutions of massless field equations in $\M_4$ with the help
of $\D$--functions introduced in \cite{Mar}. Generically,
$\D$--functions  provide the integral representation for solution
of  field equations  of the form \be \label{CD} C(X)=\int_{\Sigma}
\D(X,\XX) C(\XX) d\XX\,, \ee where $\Sigma$ is a  surface where
the initial data are given. Formulae of this type should respect a
number of properties. Firstly, $\D(X,\XX)$ should form a solution
of the field equations under consideration with respect to $X$.
Secondly, $\D(X,X^\prime)\Big |_{X, X^\prime \in
\Sigma}=\delta_\Sigma (X-X^\prime).$ Thirdly, the formula
(\ref{CD}) should be independent of local variations of $\Sigma$
which property is satisfied if $\D(X,\XX)$ solves the field
equations with respect to $\XX$ and (\ref{CD}) is defined as a
conserved charge with respect to $\XX$. The proper definition of
the integration measure $d\XX$ in (\ref{CD}) is achieved in this
paper. Note that so defined $\D$-function satisfies the
composition property \be \nn \D(X,\XX)=\int_{ \Sigma}  \D(X,\XXX)
\D(\XXX,\XX) d\XXX\,. \ee

In the analysis of HS currents, we find it most convenient to
depart from the real space $\M_M$ to its complexification $\ZIG_M$
known as upper Siegel half-space \cite{Siegel} (see also
\cite{Mumford}). It turns out that in this framework positive- and
negative-frequency solutions identify with holomorphic  and
antiholomorphic solutions in the upper Siegel half-space. $\M_M$
is a boundary (absolute) of the Siegel half-space. A surprising
conclusion will be that unfolded field equations themselves
distinguish between positive- and negative-frequency solutions of
the field equations, \ie between particles and antiparticles, the
property usually delegated to the quantization procedure. It is
worth to mention that the unfolded equations  have a form of
multidimensional Schrodinger equations. These conclusions may
eventually be of key importance for a deeper understanding of the
interplay between unfolding and quantization.

Once the HS field equations are reformulated in the Siegel space,
it is straightforward to observe that they are solved by Riemann
theta functions. This fact is truly remarkable in view of the role
that theta functions play in modern geometry, theory of integrable
systems and String Theory, and is hoped to shed more light on
fundamental structures underlying HS theory. Let us note that in
some sense theta functions provide most symmetric non-zero
solutions of massless field equations. Namely, they are invariant
up to a phase under the  transformations from the {Igusa} group
$\Gamma_{1,2}\subset Sp(2M,\mathbb{Z})$ (see \cite{Mumford} for
more detail). This class of solutions may indeed play a
distinguished role in the HS theory because the observables
constructed from such solutions, like, e.g., conserved currents,
turn out to be invariant under $\Gamma_{1,2}$. Note also that some
of the  properties of theta functions admit natural interpretation
in terms of the unfolded massless field equations.

All seemingly different aspects of HS theory considered in this
paper take their origin in the symmetry properties of the massless
field equations (\ref{dydy}), \ie HS symmetries. One of the
advantages of the unfolded formulation is just that it makes
symmetries of PDE manifest. Therefore we start in Section
\ref{unfold} by recalling some relevant facts of the unfolded
formulation approach, with the emphasize on  symmetries in
Subsection \ref{sym}. In Subsections \ref{ip} and \ref{cur} we
recall, respectively, the relationship between the ten-dimensional
space $\M_4$ and Minkowski space $M^4$ and some known results on
HS conserved currents in $\M_4$ and $M^4$. In Section
\ref{Unfolded dyn}, we extend the unfolded description of massless
higher spins to the complex Siegel space and explain how unfolded
field equations distinguish between positive- and
negative--frequency solutions. In Section \ref{Bilinear}, we
develop further the construction of conserved currents by
extending $\M_M$ to $\M_M\times \mathbb{R}^M\times \mathbb{R}^M$
where we introduce a $2M$-form, that is closed by virtue of
certain unfolded equations, and show how it reproduces the
previously known closed $M$-forms associated with conserved
currents. Then, in Section, \ref{conserv} we show how the known
$4d$ currents result from those in $\M_4$ after introducing a
singular flux in the spinning directions. Using $\D$-functions
found in \cite{Mar} and the obtained construction of conserved
currents, we derive in Section \ref{Dfunc}  the integral evolution
formulae for solutions of the massless field equations in $\M_4$.
Multilinear conserved currents are considered in Section
\ref{Multinear}. In Section \ref{thetafunc}, we show that Riemann
theta functions form a natural class of periodic solutions of
massless field equations. Conclusions and perspectives are
discussed in Section \ref{conc}. In Appendix, we describe a
commutative associative product law $\tudasuda$ which endows the
space of solutions of unfolded  equations of any rank in $\M_M$
with the commutative ring structure.

\section{Preliminaries}\label{unfold}
A natural approach to the study of dynamical equations of motion
in the HS gauge theory, referred to as {\it unfolded formulation},
consists of reformulation of PDE in the form of certain covariant
constancy conditions \cite{Ann}. Using this approach, consistent
gauge invariant nonlinear HS equations of motion were found in
\cite{PV,more,non} for HS theories in three, four and any
dimension, respectively, (see \cite{Gol,SSS,solv} for reviews and
more references). The unfolded formulation is particularly useful
for revealing symmetries and dynamical content of PDE as discussed
e.g. in \cite{tens2}. Here we briefly recall some properties of
this approach.

\subsection{Unfolded formulation and symmetries}
\label{sym} Consider a  system of linear PDE of the form \be
\label{genequ} (d +\go )C(X) = 0 \,,\qquad  d=d X^k\frac{\p}{\p
X^k}\,, \ee where $C(X) $ is a section of the trivial vector
bundle $\B= \mathbb{R}^d\times V$   over the space-time base
$\mathbb{R}^d$ with the local coordinates $X^k$ \bee\nn
\begin{array}{rcl}
V& \longrightarrow&\B\\
&&\downarrow\\
&& \mathbb{R}^d \,,
\end{array}
\eee with a linear space $V$ as the fiber. In the cases of
interest $V$ identifies with an appropriate  space of power series
$ f(Y) = \sum_{n=0}^\infty f_{\zga_1 \ldots \zga_n} (X)
Y^{\zga_1}\ldots Y^{\zga_n} $ in some auxiliary variables $Y^\za$,
\ie $C(X)$ with values in $V$ is realized as a function $C(Y|X)$
of the two types of variables.

The one-form $\go (X)=d X^k\go_k (X)$, that satisfies the flatness
condition \be \label{R0} d\go +\half [\go \,, \go ]=0\,, \ee is
some flat connection of a Lie algebra $\mathfrak{g}\subset End \,
V\,\,$ with the Lie product $[\,,]$. (Here we discard the wedge
product symbol $\wedge\,$.)

The equation (\ref{genequ}) is invariant under the global symmetry
$\mathfrak{g}$. Indeed, the system (\ref{genequ}) and (\ref{R0})
is invariant under the infinitesimal gauge transformations \bee
\nn \delta \go (X) = d \epsilon (X) + [\go (X) , \epsilon (X)]\,,
\eee \be \label{glsym} \delta C (X) = -\epsilon (X) C (X) \,, \ee
where $\epsilon (X)$ is an arbitrary symmetry parameter that takes
values in $\mathfrak{g}$. For a given $\go (X)$, there is a
leftover symmetry  with the parameter $\epsilon (X)$, that
satisfies \be \label{glpar} \delta \go (X) \equiv d \epsilon (X) +
[\go (X) , \epsilon (X)] =0\,. \ee This equation on $\epsilon(X)$
is consistent as a consequence of (\ref{R0}), \ie the Bianchi
identity $d^2=0$ does not impose any further conditions on
$\epsilon(X)$. Therefore, it reconstructs locally the dependence
of $\epsilon(X)$ on $X$ in terms of its values $\epsilon(X_0)$ at
any point $X_0$ of space-time. In the absence of topological
obstructions, the resulting global symmetry algebra with the
parameters $\epsilon(X_0)$ is $\mathfrak{g}$. It is therefore
enough to observe that some dynamical system can be reformulated
in the form (\ref{genequ}), where a flat connection $\go(X)$ takes
values in some algebra $\mathfrak{g}$ that acts in $V$, to reveal
the global symmetry $\mathfrak{g}$ (\ref{glsym}) of the system
(\ref{genequ}). This approach is general since every
$\mathfrak{g}$--invariant linear system of PDE can be reformulated
in the form (\ref{genequ}) by adding enough auxiliary variables
(nonlinear systems are described in terms of an appropriate
generalization associated with free differential algebras
\cite{FDA1,FDA2} as explained e.g. in \cite{s3,33,Ann,Gol}).

{}From this analysis it follows \cite{BHS} that the equations
(\ref{dydy}) and, hence, (\ref{oscal}) and (\ref{ofer}) are
invariant under $\mathfrak{sp}(2M,\mathbb{R})$ and its
infinite-dimensional HS extension. Indeed, the 0-form $C(Y|X)$ can
be interpreted as a section of the fiber bundle $\B=
\mathbb{R}^{\frac{M(M+1)}{2}}\times V$ where $V$ is the Fock
module of the associative Weyl algebra $A_{M}$ with the generators
$Y^A$ and $P_A$, that satisfy \be\nn [P_A\,,Y^B]=\delta_B^A\q
[Y^A\,,Y^B ] = 0\q [P_A\,,P_B ]=0\,. \ee

The  Weyl algebra  is spanned by various polynomials $a(Y,P)$. The
Fock module $V$ is spanned by the vectors
\be\nn V:\qquad
f(Y)|0\rangle
\ee generated from the Fock vacuum $|0\rangle$, that
satisfies $\dis{ P_A |0\rangle=0\,. }$ Clearly, $V$ forms a module
of $A_M$ as well as of the Lie (super)algebra
$\mathfrak{(s)hs}(2M)$ constructed from $A_M$ via
(anti)commutators with even and odd subspaces identified with the
spaces of even and odd functions $a(Y,P)$, \ie $ \nn a(-Y,-P )=
(-1)^{\pi(a)} a(Y,P)\,. $ The algebras of this type were
identified in \cite{Fort2} with the HS symmetry algebras found in
\cite{FVA}\footnote{Note that later on it was shown \cite{KV}
that, although the even subalgebra of the Weyl algebra is indeed
the  HS symmetry algebra of a nonlinear bosonic HS model with
integer spin fields, in presence of fermions both the symmetry and
the field spectrum have to be doubled by adding the Klein
operators \cite{FVau} (see \cite{33} for a recent discussion in
the context of the analysis in $\M_M$). In this paper we do not
introduce the Klein operators, which however should be  expected
in a nonlinear supersymmetric theory.}. Note that we do note rule
out the Lie algebra $\mathfrak{hs} $ resulting from $A_4$ via
commutators for all its elements because, although, having wrong
relationship between spin and statistics, it is of interest in the
context of the theory of theta functions considered in Section
\ref{thetafunc}. Note also that bosonic spinorial symmetries of
this class were recently considered in the four-dimensional setup
in \cite{FEL,FE1,BO}.

The finite-dimensional subalgebra $\mathfrak{sp}(2M\,,
\mathbb{R})\subset\, \mathfrak{(s)hs}(2M)$        is spanned by
the generators
\be\nn L_{A}{}^{B}=\half \{P_{A}\,,Y^{B}\}
\,,\qquad P_{AB}= P_{A}P_{B} \,,\qquad K^{AB}= Y^{A}Y^{B}\,. \ee

The equations (\ref{dydy}) now take the form (\ref{genequ}) with
\be \label{go0} d=dX^{AB}\f{\p}{\p X^{AB}}\q \go =\mu\, dX^{AB}
P_{AB}\,. \ee Since $[P_{AB}\,,P_{CD}]=0,$ so defined $\go$ is a
flat ${\mathfrak{sp}(2M)}$  connection, that satisfies (\ref{R0}).
Locally, any flat connection admits a pure gauge representation
\be \label{pg} \go (P,Y|X) = g^{-1} (P,Y|X)\, d\, g(P,Y|X)\,. \ee
Using this representation, we solve (\ref{glpar}) in the form \be
\label{gsy} \epsilon (P,Y|X) = g^{-1} (P,Y|X) \epsilon_0(P,Y)
g(P,Y|X)\,, \ee where $\epsilon_0(P,Y)$ is an arbitrary
$X$-independent element of $A_M$ \be \label{epar} \epsilon_0(P,Y)=
\epsilon\sum_{n,m\geq 0} \eta_{\,A_1\ldots A_n}{}^{B_1\ldots B_m}
Y^{A_1}\ldots Y^{A_n} P_{B_1}\ldots P_{B_m}\,. \ee

For the flat connection (\ref{go0}), the pure gauge representation
(\ref{pg}) holds with \be\nn g(P,Y|X) =\exp{(\mu X^{AB}P_A P_B)}\q
g^{-1} (P,Y|X) =\exp{(-\mu X^{AB}P_A P_B)}\,. \ee

\newcommand{\jj}{j}
\newcommand{\jh}{h}

Let $\epsilon_0(P,Y)$ be of the form \be\nn \epsilon_0(P,Y)
=-\gep\exp \jh^B P_B\, \,\, \exp \jj_A Y^A= -\gep\exp \jj_A\jh^A
\, \,\exp \jj_A Y^A\,\,\exp \jh^B P_B\,, \ee where $\jj_A$ and
$\jh^A$ are numerical parameters. Then the global symmetry
parameter (\ref{gsy}) takes the form \be \nn\epsilon (P,Y|X) =
-\epsilon\exp \jj_A h^A \exp{ (-\mu X^{AB}\jj_A \jj_B)} \exp \jj_A
Y^A  \, \exp ((\jh^B-2\mu X^{BC}\jj_C) P_B )\,. \ee The resulting
global symmetry transformation (\ref{glsym}) reads as \be
\label{gtr}\delta C(Y|X) = \epsilon \exp \jj_A h^A \exp{(\jj_A Y^A
-\mu  X^{AB}\jj_A \jj_B)} \, C(Y^B +\jh^B-2\mu X^{BC}\jj_C|X) \,.
\ee Note that this formula was derived in Section 7.2 of
\cite{BHS} in the Weyl ordering, \ie for $\mu=1$ and
$$
\ls\ls\ls\epsilon^\prime_0(P,Y) = \exp (\jj_A Y^A + \jh^B P_B)=
\exp \jh^B P_B \exp \jj_A Y^A  \exp(-\half \jj_A \jh^B).
$$

Differentiating (\ref{gtr}) with respect to $\jh^B$ and $\jj_A$,
it is easy to derive the transformation law  for any global HS
symmetry with polynomial parameter $\epsilon_0(P,Y)$. In
particular, at $\mu=1$ and $M=4$, the $\mathfrak{sp}(8)$
transformations are \cite{BHS} \bee \nn\label{spP}\ls\ls\ls
 && P_{AB} C(Y|X) =
\f{\p^2}{\p \jh^A \p \jh^B}  \gep^{-1}\gd C(Y|X) \Big|_{\jh^A \! =
\jj_A =0}\!= - \,\, \f{\p}{\p X^{AB}}
   \, C(Y|X)\,,
\\  \nn \ls\ls\ls\label{spL}
&&\ls\ls L_A{}^B\!C(Y|X) = \!\Big(\f{\p^2}{\p \jh^A \p\!
\jj_B}\!+\!\f{M}{2}\gd_A{}^B\!\Big) \gep^{-1} \gd C(Y|X) \Big
|_{\jh^A \! = \jj_A =0}\! = \!\Big( Y^B\! \f{\p}{\p Y^A \! }\!
+2X^{BC}\!\f{\p}{\p X^{CA}}\!+ \!\f{M}{2}\gd_A{}^B\!\Big) C(Y
|X),\quad
\\\ls\ls\ls&&   \nn\label{spK}
K^{AB}C(Y|X) =\f{\p^2}{\p \jj_A \p \jj_B}  \gep^{-1} \gd C(Y|X)
\Big |_{\jh^A \! = \jj_A =0}\!= \qquad\qquad\\ \nn\ls\ls
&&\qquad\,=\, \Big( Y^BY^A-2Y^A X^{BC} \f{\p}{\p Y^C} -2Y^B X^{AC}
\f{\p}{\p Y^C} -2X^{AB} -4X^{BC}X^{AD}\f{\p}{\p X^{CD} }\Big) \,
\, C(Y |X)\,. \eee

Note that, as explained in more detail in \cite{BHS,33},
$\mathfrak{sp}(8)$ contains  $4d$ conformal algebra
$\mathfrak{su}(2,2)$ as a subalgebra. Therefore the equations
(\ref{dydy}) are conformal invariant. Irreducible
$\mathfrak{su}(2,2)$  invariant subsystems correspond to $4d$
massless fields of different spins.

\subsection{Initial data problem}
\label{ip} Usual $d-$dimensional Minkowski space-time $\Mi$ is a
subspace of the matrix space $\M_M$ for an appropriate $M$. To
describe the embedding of $4d$ Minkowski space-time into $\M_4$ it
is convenient to use complex notations with two-component indices
$\ga,\gb $ and $\da,\db $ in place of the four-component indices $
A, B \ldots $ with the convention that the complex conjugation
interchanges unprimed indices $\ga,\gb = 1,2$ with the primed
ones $\da,\db=\done, \dtwo $. We use notation with four-component
indices being equivalent to a pair of primed and unprimed
two-component Greek indices (e.g., $ A= \ga ,\da$) and $Y^ A
=(Y^\ga  , Y^\da )$, where $ Y^\da=\overline{Y^\ga }$.

In terms of two-component complex spinors we set \bee \nn X^{ A B}
= \Big ( X^{\ga\gb} , X^{\ga \db} , X^{\da\db} \Big ) \eee so that
$X^{\da\db} $ is complex conjugated to $X^{\ga\gb}$, \ie
$\overline{X^{\ga\gb}}=X^{\da\db}$, while  $X^{\ga \db}$ is
hermitian, $\overline{X^{\ga\db}}= X^{\gb\da}$. For Minkowski
coordinates, we will sometimes use notation $x^{\ga \db}$ instead
of $X^{\ga \db}$. Two-component indices are  raised and lowered
according to \bee \nn A^\ga=\gvep^{\ga \gb}A_\gb\q
A_\gb=\gvep_{\ga \gb}A^\ga\q \gvep_{\ga\gb} = - \gvep_{\gb\ga}\q
\gvep_{12} = 1\,, \eee and analogously for primed indices.

Minkowski time $t$ and space coordinates $x^i$ are \bee
\label{4dfib} X^{\ga \db} =t\T^{\ga\db} +x^i \sigma_i^{\ga \db}\q
i=1,2,3 \,,\eee where $\T^{\ga\db}=\delta^{\ga\db}$ while
$\sigma_i^{\ga \db}$ are hermitian traceless Pauli matrices. The
Klein-Gordon and Dirac equations in Minkowski space read as \be
\label{mineq} \Big ( \f{\p^2}{\p X^{\ga \db} \p X^{\gga\dd}} -
\f{\p^2}{\p X^{\ga \dd} \p X^{\gga\db}}\Big ) b(X) =0\,, \ee \be
\label{Dir1} \f{\p}{\p X^{\ga \db}} f_\gd (X) - \f{\p}{\p X^{\gd
\db}} f_\ga (X)      =0\q \f{\p}{\p X^{\ga \db}} f_\dga(X)       -
\f{\p}{\p X^{\ga \dga}} f_\db(X)      =0\,. \ee

As  shown in  \cite{Mar,s3}, the generalized space-time $\M_M$
admits the well-defined notions of future and past. The
(past)future cones of the origin $X=0$ are formed by
(negative)positive-definite matrices $X^{ A B}$. There is a single
time evolution parameter \be \label{t} t=\f{1}{M} X^{ A B} \T_{ A
B}\,, \ee where $\T^{ A B}$ is some positive-definite time-arrow
matrix. For a chosen time parameter $t$, the global space-like
Cauchy surface $\Sigma_t$ is parameterized as \be  \label{res} X^{
A B} \in \Sigma_t :\qquad X^{ A B} = x^{ A B}+t\T^{ A B}\,, \ee
where the space coordinates $x^{ A B}$ are arbitrary $\T-$
traceless matrices \be \label{xtr} x^{ A B}\T_{ A B}=0\,,\qquad
\T_{ A B}\T^{ B C} =\delta_ A^ C\,. \ee

A particular solution in   $\M_M$ can be reconstructed from the
values of the fields along with some their time derivatives on the
global Cauchy surface. However, because the system of equations
(\ref{oscal}) on the field $b(X)$ is overdetermined, some of these
equations play a role of constraints that restrict the choice of
the initial data on the global Cauchy surface. Independent initial
data can be given on a lower-dimensional object called local
Cauchy bundle $E$, which is a $M-$dimensional fiber bundle over a
$(d-1)$-dimensional base manifold $\sigma \in \Sigma $ treated as
the space manifold. The Minkowski space-time is $R \times \sigma
\subset \M_M$ where $R$ is a time axis.

To see that  initial data for the equations (\ref{oscal}),
(\ref{ofer}) should be given on a $M$-dimensional surface is most
convenient by using their unfolded form (\ref{dydy}). Indeed, the
generic solution of   (\ref{dydy}) can locally be given in the
form \bee\nn C(Y|X) =\exp\left(-\mu X^{ A B}\f{\p^2}{\p Y^ A \p Y^
B }\right)C(Y|0)\,, \eee where the ``initial data'' $C(Y|0)$ is an
arbitrary function of $M$ variables $Y^ A$.

Note that the equivalence of the unfolded equations (\ref{dydy})
to (\ref{oscal}), (\ref{ofer}) shown in \cite{BHS} manifests
itself in the inverse formula that reconstructs the dependence of
$C(Y|X)$ on $Y$ in terms of any functions $b(X)$ and $f_A (X)$
that satisfy (\ref{oscal}) and (\ref{ofer}), respectively,
\bee\label{inv} \ls C(Y|X) = \cos (s)b(X)+ \f{\sin(s)}{s}Y^A
f_A(X)\q s=\sqrt{\f{1}{\mu} Y^A Y^B \f{\p}{\p X^{AB}}}\,. \eee

\subsection{Higher-spin currents}
\label{cur} The infinite set of conformal HS symmetries found in
\cite{BHS} is parameterized by various global symmetry parameters
(\ref{epar}). This suggests the existence of the corresponding
conserved HS currents. Indeed, in \cite{cur} it was shown that the
$M$-form in $\M_M$ \bee \label{clo} \Omega_M(\eta, C^k, C^l) &=&
\epsilon_{ C_1 \ldots  C_M} dX^{ C_1 A_1}\wedge \ldots \wedge dX^{
C_M  A_M}\qquad\\ \nn &{}& \eta_{ B_1 \ldots B_n}{}^{ A_{M+1}
\ldots  A_{M+m}} X^{ A_{M+m+1}  B_1}\ldots X^{ A_{M+m+n} B_n}
T^{kl}_{ A_1 \ldots  A_{M+m+n}}(X)\,, \eee where $\epsilon_{ C_1
\ldots C_M}$ is the totally antisymmetric multispinor and
constants $\eta$ are the HS symmetry parameters, is closed
provided that the generalized stress tensor $T^{kl}_{ A_1 \ldots
A_{N}}(X)$ \be \label{stress2} T^{k\,l} {}_{ A_1 \ldots
A_{N}}(X)\,= \f{\p}{\p Y^{ A_1}}\dots \f{\p}{\p Y^{ A_N}} C^k
(Y|X) C^l (i Y|X)\,\Big|_{Y=0} \ee is built of the fields $ C^k
(Y|X)$ and $C^l (i Y|X) $ that satisfy  (\ref{dydy}). (Here $k$
and $l$ are color indices which take an arbitrary number of
values.) The charge \be \label{Q4} Q(\eta, C^k, C^l) =
\int\limits_{E^M} \Omega_{M}(\eta, C^k, C^l) \,, \ee is
independent of  local variations of a $M$-dimensional surface
$E^M$, \ie it conserves.

On the other hand,  HS charges in Minkowski space have the form
\be \label{Q3} Q(\eta, C^k, C^l) = \int\limits_{\sigma^{d-1}}
\Omega_{d-1}(\eta, C^k, C^l) \,, \ee where $\Omega_{d-1}(\eta,
C^k, C^l) $ is a on-mass-shell closed $(d-1)$-form dual to the
conserved current, and $\sigma^{d-1}$ is a $(d-1)$-dimensional
surface in the Minkowski space-time, usually identified with the
space surface $R^{d-1}$. The  explicit expression for the on-shell
closed three-form $\Omega_{3}(\eta, C^k, C^l)$ in $4d$ Minkowski
space, obtained recently in \cite{GSV}, is
$$
\Omega_3(\eta, C^k, C^l) = dx_{\ga\da}\wedge
dx^{\ga{\gga}^\prime}\wedge dx^{\gga\da} w_\gga{w}_{\dga}
\eta({w,u}) C^k(Y|x)C^l(iY|x) \Big |_{Y=0}\,,
$$
where
$$
w_\ga =\frac{\partial}{\partial Y^\ga},\,\,\,
\bw_\da=\frac{\partial}{\partial Y^{\da}},\,\,\, u^\ga
=x^{\ga\da}\frac{\partial}{\partial Y^{\da}},\,\,\,
\bu^{\da}=x^{\ga\da} \frac{\partial}{\partial Y^\ga}\,.
$$
Equivalently\footnote{We use the convention where a number in
parentheses next to an index denotes a number of symmetrized
indices. For instance, $\ga(n)$ stands for $n$ symmetrized indices
$\ga_1\ldots \ga_n$}, \bee \label{clo3} \Omega_3(\eta, C^k, C^l)=
\nn dx_{\eta \eta^\prime} \wedge dx^{\eta \dga} \wedge
dx^{\gga\eta^\prime} \,\\
\eta_{\ga(n)\,\da(m)}{}^ {\gb(p)\,\,\db(q)} x^{\ga _{1}
\dm_1}\ldots x^{\ga _{n}\dm_n} x^{\gm _{1} \da_1}\ldots x^{\gm
_{m}\da_m} T^{k\,l}_{\gga \, \gb(p)\,\,\gm(m)\,\,\dga\,\db(q)\,\,
\dm(n)\,\,} (x) \eee is closed provided that $ T^{k\,l}_{ A_1
\ldots  A_{n}}(x) $ is the restriction of the stress tensor
(\ref{stress2}) to the Minkowski space $M^4\subset \M_4$, that has
only nonzero coordinates $X^{\ga\db}=x^{\ga\db}$ among $X^{AB}$,
and  the fields $ C^k (Y|x)$ satisfy the $4d$ unfolded equations
\bee \label{minun} \f{\p}{\p x^{\ga \db}}C^k(Y|x) +\f{\p^2}{\p
Y^\ga \p Y{}^{\db}}C^k(Y|x)=0\,. \eee Let us note that the
equations of this type naturally appear in the study of so-called
twistorial world-like particle models (see e.g. \cite{shir,BC}) as
the Fourier transformation of the Dirac constraints on the
respective phase space momenta.

It was conjectured in \cite{cur} that the charge (\ref{Q4}) at
$M=4$ should reproduce the Minkowski HS  charge via an appropriate
reduction of $\M_4$ to  $M^4$. This conjecture sounds plausible
because the two  charges contain the same symmetry parameters and
the Minkowski stress tensor results from the restriction  of the
generalized stress tensor in $\M_4$ to the Minkowski space. One of
the  goals of this paper is to establish the precise
correspondence between the $\M_4$ and Minkowski realizations of
the conserved charges.

For the case of $M=2$ the identification is a sort of trivial
because $M^{3} = \M_2$ and the local Cauchy bundles are the same,
both being two-dimensional. (Note that since $3d$ conformal
Minkowski charges are constructed from $3d$ massless scalar and
spinor they coincide with the $3d$ conformal HS currents found in
\cite{KVZ}.) For higher dimensions the precise identification is
less trivial. The problem is that the dimensions of local Cauchy
bundles are different in $\Mi$ and $\M_M$ for $M >2$. The extra
dimensions in $E^M$ compared to $\sigma^{d-1}$ are responsible for
spin and are associated with the corresponding compact spaces. It
turns out  \cite{BLS,Mar,BBAST} that $E^M = \mathbb{R}^{d-1}\times
S^{M-d+1}$ for $M=2,4,8,16$ correspond to $d=3,\,4,\,6,\,10$. In
particular, the $\,M=4\,$ spin space  is $S^1$.

\bigskip
\section{Quantization and Siegel space}
\label{Unfolded dyn}

The coefficient $\mu$ in front of the second term in the unfolded
equations (\ref{dydy}) was irrelevant within an expansion in
powers of $Y^\A$  used in \cite{BHS}. Its absolute value can
certainly be normalized arbitrarily by a rescaling of $Y^A$.
However, its phase should respect reality conditions.
Surprisingly, it  distinguishes between positive and negative
frequencies, \ie  particles and antiparticles. We interpret this
observation as an indication that the unfolded dynamics  encodes
quantum physics.

Indeed, general solution of the equations (\ref{oscal}) and
(\ref{ofer}) is \cite{Mar} \bee \label{bfo} b (X) \,=\,b^+ (X)
\,&+&\,b^- (X) \,=\\\nn\,\f{1}{\pi^{\f{M}{2}}} \int d^M\xi\, \Big
( b^+ (\xi ) \exp\big\{ i  \xi_{\za}\xi_{\zb} X^{\za\zb}\big\}
&+&b^- (\xi ) \exp\big\{ -i  \xi_{\za}\xi_{\zb}X^{\za\zb}
\big\}\Big ) \, \eee and \bee \label{ffo} f_\zga (X)
\,=\,f_\zga{}^+ (X) \,&+&\,f_\zga{}^- (X) \,=\\\nn
\f{1}{\pi^{\f{M}{2}}} \int d^M\xi\,\,\,\,\, \xi_\zga \Big (  f^+
(\xi ) \exp \big\{i  \xi_{\za}\xi_{\zb} X^{\za\zb}\big\} &+& f^-
(\xi ) \exp\big\{ -i  \xi_{\za}\xi_{\zb}X^{\za\zb}\big\} \Big )
\,. \eee Note that $b^\pm(\gl)=(-1)^{M}b^\pm(-\gl)$ and
$f^\pm(\gl)=(-1)^{M+1}f^\pm(-\gl)$. In particular, for even $M$,
$b^\pm (\xi)$ and $f^\pm(\xi)$ are even and odd functions of
$\xi$, respectively. The integration in (\ref{bfo}) and
(\ref{ffo}) is hence over $\mathbb{R}^M /  {Z}_2$. The point
$\xi_\ga =0$ is invariant under the ${Z}_2$ reflection $\xi_\ga\to
-\xi_\ga$ and therefore is a singular point of the orbifold
$\mathbb{R}^M /  {Z}_2$.

Now we observe that the unfolded equations \be \label{dydyh+}
\left ( \f{\p}{\p X^{ A B}} \pm \,i\,\hbar \f{\p^2}{\p Y^ A \p Y^
B}\right ) C^\pm(Y|X) =0\, \ee distinguish between the positive--
and negative--frequency solutions \be \label{cbfopmX+} C^\pm (Y|X)
=\f{1}{\pi^{\f{M}{2}}} \int d^M\xi\,c^\pm (\xi) \exp \pm  \,i \Big
( \hbar\,\, \xi_{\za}\xi_{\zb} X^{\za\zb}+Y^B \gx_B \Big )\,, \ee
that are complex conjugated to each other for real $X$ and $Y$,
\bee\label{cc} c^- (\xi) = \overline{c^+ (\xi)}\q C^- (Y|X) =
\overline{C^+ (Y|X)}\,. \eee Note also that, for even~$M$,
$c^\pm(\xi)$ contain $b^\pm (\xi)$ and $f^\pm (\xi)$ as  even and
odd parts, respectively, \bee \label{decomio}
c^\pm(\xi)=b^\pm(\xi) + f^\pm (\xi)\,. \eee

As explained in more detail in \cite{Mar}, the manifest
decomposition into positive-- and negative--frequency parts  gives
rise to the quantum fields with the creation and annihilation
operators $\hat{c}^\pm(\xi)$, that satisfy the commutation
relations \be \label{qpr} [\hat{c}^\pm (\xi_1 ) , \hat{c}^\pm
(\xi_2 )] =0\,,\qquad [\hat{c}^- (\xi_1 ) , \hat{c}^+ (\xi_2 )] =
\delta (\xi_1 - \xi_2 )\,. \ee In this paper we will be mostly
interested in the classical picture, however.

For the further analysis it is convenient to introduce   complex
coordinates \be\label{Sieg} \Z^{AB}
=X^{\za\zb}+i\,\X^{\za\zb}\,\,\equiv \Re \Z^{AB}+i \Im \Z^{AB}.
\ee The real part $\Re \Z^{AB}$  of $\Z^{\za\zb}$ is identified
with the coordinates of the generalized space-time $X^{\za\zb}$
that contain in particular Minkowski coordinates. The imaginary
part $\X^{\za\zb}=\Im\Z^{\za\zb}$ is required to be positive
definite and was treated in \cite{Mar} as a regulator that makes
the Gaussian integrals well-defined (\ie physical quantities are
obtained in the limit $\X^{\za\zb}\to 0$; note, that the complex
coordinates $Z^{\za\zb}$ introduced in \cite{Mar} are related to
$\Z^{\za\zb}$ as $ \Z^{AB} = i\bZ^{AB}$). The space  of
coordinates $\Z^{\za\zb}$ forms the upper Siegel half-space
$\ZIGM$ \cite{Siegel,Mumford}. Evidently,  $-\cZ^{AB}\in \ZIGM $
provided that $\Z^{AB}\in \ZIGM $ and vice versa.

The variables $Y^A$ can also be  complexified
$$\gY^{\za}=Y^{\za}+i\,\Y^{\za},$$
extending the  Siegel space to \emph{ Fock-Siegel space} $\GZM$.

The continuations of the functions $C^\pm$(\ref{cbfopmX+}) to  the
Fock-Siegel space $\GZM$ are \bee \label{Csieg+} C^+(\gY|\Z)
&=&\f{1}{\pi^{\f{M}{2}}}\int d^M \gx\,\,c^+(\gx)\exp
\big(i\,\hbar \gx_A\gx_B \Z^{AB}\big)\,
 \exp i\big( \gx_A \gY^A\big)\,,
\\
 \label{Csieg-} C^-(\cY|\cZ)
 &=&\f{1}{\pi^{\f{M}{2}}}\int d^M \gx\,c^-(\gx)\exp \big(- i\,\hbar \gx_A\gx_B \cZ{}^{AB} \big)\,\exp - i\big(  \gx_A \cY{}^A\big),
\eee where $c^\pm(\xi)$ are some ``Fourier coefficients".
Depending on a problem in question, they can be chosen to belong
to different functional classes.

The broadest framework   is provided by   distributions that grow
not faster than exponentially of order two and zero type at
infinity, \ie not faster than $\exp A|\gx|^2$, $\forall A>0$. In
this case, $c^\pm(\xi)$ belong to the space $S^{\,\prime}_{1/2}
(\mathbb{R}^M)$ dual to the Gelfand-Shilov space $
S_{1/2}(\mathbb{R}^M) $~ \footnote{ Recall that $
S_{\ga_1,\dots,\ga_M} $ is defined in \cite{GelShil} as a space of
infinitely differentiable functions $\phi(x_1,..,x_M)$ such that
the inequality $ \Big|x_1^{k_1}\dots x_M^{k_M}
\,\f{\p^{q_1+\dots+q_{M}}}{\p x_1^{q_1}\dots \p
x_M^{q_M}}\,\phi(x)\Big| \le C_q A_1^{k_1}\dots
A_M^{k_M}\,k_1^{k_1 \ga_1}\dots k_M^{k_M \ga_M} $ holds for any
integer nonnegative $k_i $, $q_i $ and some constants $C_q$ and
$A_i$ that depend on $\phi$. The Fourier dual space
$S^{\ga_1,\dots,\ga_M}$ consists of infinitely differentiable
functions $\psi(p_1,..,p_M)$ that satisfy  the inequality $
\Big|p_1^{k_1}\dots p_M^{k_M} \, \f{\p^{q_1+\dots+q_{M}}}{\p
p_1^{q_1}\dots \p p_M^{q_M}}\,\psi(p) \Big| \le C_k B_1^{q_1}\dots
B_M^{q_M}\,q_1^{q_1 \ga_1}\dots q_M^{q_M \ga_M} $ for any integer
nonnegative $k_i $, $q_i $ and some constants $C_k$ and $B_i$ that
depend on $\psi$.
\qquad $S_{1/2}(\mathbb{R}^M)$\, and \,$S^{1/2}(\mathbb{R}^M)$ are \,shorthand \, notations for 
$S_{{\underbrace{\mbox{\tiny 1/2,\dots,1/2}}_{\mbox{\tiny M}}}}
\mbox{\quad and\quad} S^{{\overbrace{\mbox{\tiny
1/2,\dots,1/2}}^{\mbox{\tiny M}}}}$, respectively . }. It can be
shown \footnote{We are grateful to M.A. Soloviev for communicating
to us this fact.}, that $C^+(\gY|\Z)$ (\ref{Csieg+}) and
$C^-(\cY|\cZ)$ (\ref{Csieg-}) are, respectively, analytic and
antianalytic in $\gY$. The  (anti)analyticity of
\Big($C^-(\cY|\cZ)$  \Big) $C^+(\gY|\Z)$ in  $\Z \in \ZIGM $
follows from the integral representations \Big((\ref{Csieg-})\Big)
(\ref{Csieg+}).

As a subclass, one can require $c^\pm(\xi)$ to be infinitely
differentiable functions, that grow not faster than exponentially
of order two and zero type at infinity. Then, for any $\Z\in
\ZIGM$ the functions $\dis{c^+(\gx)\exp  \big(i\,\hbar \gx_A\gx_B
\Z^{AB}\big)\, }$   and $\dis{c^-(\gx)\exp \big(- i\,\hbar
\gx_A\gx_B \cZ{}^{AB} \big)\,}$ belong to $ S_{1/2}(\mathbb{R}^M)$
with respect of $\gx$ and hence  their Fourier images
$C^+(\gY|\Z)$ (\ref{Csieg+})   and $C^-(\cY|\cZ)$ (\ref{Csieg-})
belong to $ S^{1/2}(\mathbb{R}^M)$. It is worth to note that, as
shown in \cite{Solov,Solov1}, the class $ S^{1/2}(\mathbb{R}^M)$
plays a distinguished role in the analysis of convergency of power
series in the Moyal star--product in noncommutative field theory.
This is particularly interesting taking into account that the
interactions of HS fields (for more detail on the role of
star--product in HS theories see \cite{BHS,more,non} and reviews
\cite{Gol,SSS,solv} ) is governed by the Moyal star--product,
which however acts on the noncommutative spinor variables  $Y^A$
rather than on the space--time coordinates as in noncommutative
field theory.

Further restrictions may be imposed in the case where the fields
have to be normalizable with respect to one or another norm. This
is  needed to guarantee that the bilinear currents are
well--defined. As discussed in more detail in Subsection
\ref{class}, the relevant classes of functions $c^\pm(\xi)$
include \,Sobolev spaces\, $L_2^q(\mathbb{R}^M)$ and
 Schwartz space  $S(\mathbb{R}^M)$.

It is easy to see, that $C^+(\gY|\Z)$ and $C^-(\cY|\cZ)$ are
complex conjugated as a consequence of (\ref{cc}), \ie $
\overline{C^+(\gY|\Z)}=C^- (\cY|\cZ)\, $ and \be \label{dgydgyh+}
\left ( \f{\p}{\p \Z^{ A B}} + \,i\,\hbar \f{\p^2}{\p \gY^ A \p
\gY^ B}\right ) C^+(\gY|\Z) =0\,, \ee \be \label{dgydgyh-} \left (
\f{\p}{\p \cZ^{ A B}} - \,i\,\hbar \f{\p^2}{\p \cY^ A \p \cY^
B}\right ) C^-(\cY|\cZ) =0\,. \ee The equations  (\ref{dgydgyh+})
and (\ref{dgydgyh-}) uplift the massless field equations for
(negative)positive frequencies to the
 Fock-Siegel space.
The (anti)holomorphy properties of   $C^\pm $  reconstruct them in
the  Fock-Siegel space in terms of theirs boundary values
$C^\pm(Y|X)$ at $\M_M\times\mathbb{R}^M$. Remarkably, depending on
the sign of the second term, the classical unfolded  field
equations (\ref{dgydgyh+})  and (\ref{dgydgyh-}) distinguish
between positive and negative frequencies, the property usually
delegated to a quantization prescription. Let us note that this
phenomenon also takes place in Minkowski setup with (appropriately
complexified) unfolded equations (\ref{minun}) as well as in the
related twistorial world-line particle models \cite{shir,BC}. In this case,
the analogues of the upper and lower Siegel spaces are the forward and
backward  tubes.

The fields $C^\pm$ can be unified into the field
\be
\label{decomp} C(\gY,\cY|\Z,\cZ) = C^+ (\gY|\Z) + C^- (\cY|\cZ)\,,
\ee that, however, does not possesses  definite (anti)holomorphy
properties in $\Z^{AB}$.

\section{Bilinear currents}
\label{Bilinear}

\subsection{Current equations}

\renewcommand{\W}{W}
\renewcommand{\U}{U}

Let us introduce a conserved current which generalizes that of
\cite{cur} in a way convenient for the further analysis. The key
fact is that a differential $2M$--form \bee\label{varpi}
\varpi^{2M}(g)\,=\, \,\left(d\,\W_A \wedge \Big(\hbar\,\W{}_B{}
d\, \Z{}^{AB} -    d\,\gY{}^A
\Big)\right)^{\,M}\,\,g(\W,\,\gY\,|\Z) \eee is closed in a domain
in $\mathbb{C}^{\f{M(M+1)}{2}}(\Z^{AB})\times
\mathbb{R}^{M}(\W_B)\times  \mathbb{C}^{M}(\gY^A) $ provided that
$g(\W,\,\gY\,|\Z)$ is holomorphic in the variables $\gY\,$ and $ \Z$ 
and satisfies the following {\it current} equations \be
\label{unfol2_Fur} \left( \, \f{\p}{\p \Z{}^{AB}}  + \,\hbar\,
\W_{(A} \f{\p}{\p   \gY{}{\,}{}^{B)}} \right)
g(\W,\,\gY\,|\Z)=0\,. \ee

Indeed, from  (\ref{varpi}) and (\ref{unfol2_Fur}) it follows that
\bee\nn &&\left (d\,\W_{A} \f{\p}{\p \W_{A}}+ d\,\Z{}^{AB}
\f{\p}{\p \Z{}^{AB}} +  d\,\gY^{A} \f{\p}{\p \gY^{A}}\right
)\wedge \varpi^{2M}(g(\W,\,\gY\,|\Z ))\,=
\\ \nn
= &&\left ( d\,\W_{A} \f{\p}{\p \W_{A}} \,- \,\Big(\hbar\,\W{}_B{}
d\,\Z{}^{AB} -d\,\gY{}^A \Big ) \f{\p}{\p \gY^{A}}\right)\wedge
\varpi^{2M}(g(\W,\,\gY\,|\Z))=0 \eee because \bee \nn d\W_C \wedge
\left(d\,\W_A \wedge \Big(\hbar\,\W{}_B{} d\,\Z{}^{AB} -d\,\gY{}^A
\Big)\right)^{\,M} =0\, \eee and \bee \nn \Big(\hbar\,\W{}_D{}
d\,\Z{}^{CD} -d\,\gY{}^C \Big )\wedge \left(d\,\W_A \wedge
\Big(\hbar\,\W{}_B{} d\,\Z{}^{AB} -d\,\gY{}^A \Big)\right)^{\,M}
=0\,. \eee

As a result, on solutions of (\ref{unfol2_Fur}), the charge \be
\label{Q} Q= Q(g)=\int_{\Sigma^{2M}} \varpi^{2M}(g) \ee is
independent of local variations of a $2M$-dimensional integration
surface $\Sigma^{2M}$. In particular, for functions that {decrease
fast enough at space infinity}, it is independent of the time
parameter in $\M_M$, thus being conserved.

Since  (\ref{unfol2_Fur}) is a first-order PDE system, the space
of its regular solutions forms a commutative algebra $\R$, \ie a
linear combination of products of any regular solutions of
(\ref{unfol2_Fur}) is also a solution.  The algebra $\R$ is formed
by functions $\eta$ of the form \be \label{par}
\eta(\W,\,\gY\,|\Z) = \varepsilon(\W_A,\,\gY^C
\,-\hbar\,\Z{}^{CB}\,\W_B)\, \ee
with arbitrary regular  $\varepsilon(\W,\gY)$. An extension of
this property to the space of singular solutions $\S$ is that $\S$
forms an $\R$-module, \ie although it may not be possible to
multiply singular solutions with themselves, they can by
multiplied by  regular ones.

To make contact with the currents (\ref{clo}) note, that
Eq.~(\ref{varpi}) gives rise to conserved currents for
$g(\W,\,\gY\,|\Z) $ of the form \be\label{eta_f} g(\W,\,\gY\,|\Z)
= \eta(\W,\,\gY\,|\Z) f(\W,\,\gY\,|\Z)\,, \ee where
$\eta(\W,\,\gY\,|\Z)$  (\ref{par}) is a polynomial solution of
(\ref{unfol2_Fur}) identified with a HS symmetry parameter and
\bee
\label{Fourier} f ( \W,\,\gY\,,|\Z)\,\,=\,\, 
(2\pi)^{-M/2}\int\limits_{\mathbb{R}^M}  
d^{\,M}\U\,\,\, \exp\left(-i\,\,\,
\W_C\,\U{}^C\right)\,T(\U,\,\gY\,|\Z)\, \eee is a  solution built
of massless fields via the generalized stress tensor
\be\label{Tbilin} T(\U,\,\gY\,|\Z) =C^+( \gY-\U|\Z)\,C^-(
{\U}+\gY|\Z), \ee where $C^+(  \gY|\Z)$ satisfies (\ref{dgydgyh+})
while $C^-( \cY|\cZ)$ satisfies (\ref{dgydgyh-}). (\,For more
accurate definition that respects  necessary analyticity
properties see Subsection \ref{stri}. The appropriate classes of
functions $C^+(  \gY|\Z)$ and $C^-(\cY|\cZ)$ will be specified in
Subsection \ref{class}.\,\,) The inverse transform is \bee
\label{fourierH}
T (\U,\,\gY\,|\Z)= (2\pi)^{-M/2} \int\limits_{\mathbb{R}^M}   
d^{\,M}\,\W\,\,\, \exp\left(i\,\,\, \W{}_C\,\U^C\right)\, f
(\W,\,\,\gY\,|\Z)\,. \eee
The equations (\ref{unfol2_Fur}) for $f(\W,\,\,\gY\,|\Z)$
translate to the following equations for the stress tensor \be
\label{unfol2uy}\left\{ \,\f{\p}{\p \Z{}^{AB}}  - i \hbar \,\,
\f{\p}{\p \gY^{(A}} \f{\p}{\p \U^{B)}} \right\} T(\U,\,\gY|\Z) =0,
\ee \ie $f(\W,\,\gY\,| \Z)$ satisfies (\ref{unfol2_Fur}) provided
that $T(\U,\,\gY\,| \Z)$ satisfies (\ref{unfol2uy}) and vice
versa. One can make sure, that the bilinear tensor (\ref{Tbilin})
satisfies (\ref{unfol2uy}) by virtue of (\ref{dgydgyh+}),
(\ref{dgydgyh-}).

Note that, up to a factor of $i \hbar$,  the equations (\ref{unfol2uy}) at real $X=\Re\Z$, $Y=\Re\gY$ 
\be \label{unfol2uyR}\left\{ \,\f{\p}{\p X{}^{AB}}  - i \hbar \,\,
\f{\p}{\p Y^{(A}} \f{\p}{\p U^{B)}} \right\} T(U,\,Y| X) =0, \ee
coincide   with the rank$-2$ unfolded equations of \cite{tens2}.
In particular, (\ref{unfol2uyR}) implies \bee  \label{okno} \ls
\left(\! \f{\p^3}{\p U^{A}\p U^{B}\p X^{CD}}\!+\! \f{\p^3}{\p
U^{C}\p U^{D}\p X^{AB}} \!-\! \!\f{\p^3}{\p U^{C}\p U^{B}\p
X^{AD}}\!-\! \f{\p^3}{\p U^{A}\p U^{D}\p X^{CB}}
\!\right)\!T(U,Y|X)\!=\!0,\rule{15pt}{0pt} \eee which equation was
used in \cite{cur} to prove that the form  (\ref{clo}) is closed.

\renewcommand{\U}{\mathcal{U}}
\renewcommand{\W}{\mathcal{W}}
\subsection{Siegel strip and bilinear currents }
\label{stri}

To define the integration around singularities in Section
\ref{conserv} via a deformation of the integration surface over
$X^{AB}$ in $\M_4$  to the complex space, we now introduce a
generalized stress tensor that depends on positive- and
negative-frequency solutions $C^+(\gY|\Z)$ and $C^-(\cY|\cZ)$ of
the unfolded equations  (\ref{dgydgyh+}) and (\ref{dgydgyh-}),
respectively, and possesses proper holomorphy properties in
$\Ds_M^{(\Hh)}(\Z)\times \mathbb{C}^{M}(\gY)$, where  a domain
$\Ds_M^{(\Hh)}\subset\ZIGM$ will be specified below. In the rest
of this section we set $\hbar=1$.

For a positive definite symmetric matrix $\Hh^{AB}$ we introduce
Siegel $\Hh$--strip $\Ds_M^{(\Hh)}\subset\ZIGM\,$  as follows :
\bee\nn
\Z^{AB}\in \Ds_M^{(\Hh)}\,&:\,&\left\{ \beee{r } (\Hh-\Im
\Z)^{AB}\quad \mbox{is positive definite}\\ \nn \Im \Z^{AB}\quad
\mbox{is positive definite} \eeee \right. \eee which is a
generalization of a strip $0< \Im z< \Hh$ in $ \mathbb{C}$ to the
Siegel space $\ZIGM$.

\medskip

Since $C^-(\cY|\cZ)$ is anti-holomorphic in $ \Ds_M^{(\Hh)}\times
\mathbb{C}^{M}$, one can see that \bee\label{C_H}
C^-_\Hh{}(\gY|\Z)=C^-(\gY|\Z-i\Hh)\eee is holomorphic in
$\Ds_M^{(\Hh)}\times \mathbb{C}^{M}$\,.
Note, that $\Hh$ is a parameter of  $C^-_\Hh{}$ and
  $$
\lim_{\Hh\to0}C^-_\Hh{}( Y|X)=C{}^-( Y|X) \qquad     \forall
\,\,\, X \in \M_M\,,\,\,\,Y \in \mathbb{R}^M.
$$

It is easy to see that a  generalized stress tensor
\bee\label{T_H} T_\Hh{} (\U,\,\gY|\Z)=
C^+(\gY-\U|\Z)\,C^-_\Hh{}(\gY+\U|\Z) \eee solves (\ref{unfol2uy})
and is   holomorphic in  $\Ds_M^{(\Hh)}\times \mathbb{C}^{2M}$. Up
to a linear change of variables, its restriction to the real
subspace in the limit $\Hh\to0$ gives the generalized stress
tensor of \cite{cur}. To reproduce the charge (\ref{Q4}) we
proceed as follows. Let the  integration surface be \bee\nn
\mathbf{\Sigma}^{2M}_{\Hh}=\sigma^M_{\Hh}(X)\times
\mathbb{R}^M(W)\Big |_{ \{\gY=\gY_0\}}\,, \eee where
\,\,$\sigma^M_{\Hh}(X)$ is any $M-$dimensional surface that
belongs to the real subspace of $\Ds_M^{(\Hh)}$ defined by the
equation $\{\,\Im \Z= \gn\Hh\}$, where $0<\gn<1$ is a free
parameter. \quad Substituting the $2M$-form $ \varpi^{2M}(g)\,$
(\ref{varpi}) with the function $g$ (\ref{eta_f}) into (\ref{Q}),
we obtain \bee \label{oldnew1} Q{}_{\Hh} \sim\!\!
\int\limits_{\mathbf{\Sigma}^{2M}_{\Hh}} \left(d\,W_A \wedge\,
W{}_B{} d\,X^{AB} \,\right)^{M}\, \eta(W,\,\gY_0\,|X+ i\gn\Hh)
f(W,\,\gY_0\,|X+ i\gn\Hh)\,,\qquad \eee whence, using the Fourier
transform (\ref{Fourier}), we get in the limit $\Hh\to 0$ \bee
\ls\ls Q_0\sim  \nn \int\limits_{\mathbf{\Sigma}^{2M}_0} \nn
\left(d\,W_A \wedge  W{}_B{} d\,X^{AB}\right)^{M}
\int\limits_{\mathbb{R}^M} d^{M}U\,  \exp\left(-i\, W{}_C
U^C\right) \, \eta(W,\,\gY_0\,|X ) T_0(U,\gY_0|X)\,
\,\sim\quad \\ \ls\ls\ls\ls\ls\label{oldnew}\\\nn %
\int\limits_{\sigma^{M}_0}\!\gep_{A_1...\A_M} d\,X^{A_1 B_1}\!
\wedge ...\wedge  d\,X^{A_M B_M}\,\, \eta\left(-i\f{\p}{\p\,U}
,\,\gY_0\,|\, X\right) \f{\p}{\p\,U{}^{B_1} }\dots
\f{\p}{\p\,U{}^{B_M} }\,\, T(U,\gY_0|X )\Big|_{U=0}\,. \quad \eee
For   the parameter $\eta$ of the form (\ref{par}) with polynomial
$\gvep$ and $\gY_0=0$ this gives  (\ref{Q4}).

\bigskip

Alternatively, one can integrate over $d^M \gY$ at fixed $\Z$. For
example let $\mathbf{\Sigma}^{2M}=\mathbb{R}^{2M}:\{ \Z=X_0 ,
\,\,\,\gY=Y \}$ for some real $X_0$. We have in the limit $\Hh\to
0$ \bee \nn \ls Q_0 \sim \int\limits_{\mathbb{R}^{2M}} \nn
\left(d\,W_A \wedge d\,Y{}^A
\right)^{\,M}\int\limits_{\mathbb{R}^M}
d^{\,M}\,U\, 
\exp\left(-i\,\,\, W{}_C\,U^C\right)\, \varepsilon\left(W ,\,Y^C
\,-\,X_0{}^{CB}\,W_B\right)\, T_0(U,\,Y\,|\, X_0)\,\sim \eee\bee
\!\!\label{d4Y}   \sim
 \int\limits_{\mathbb{R}^M}
\gep_{A_1\dots\A_M} \, d\,Y^{A_1} \,\wedge \dots\wedge  d\,Y^{A_M
} \varepsilon\left(-i\f{\p}{\p\,U}  ,\,Y^C
\,+i\,X_0{}^{CB}\,\f{\p}{\p\,U^B} \right)\, T_0(U,\,Y\,|
\,X_0)\Big|_{\,U=0}\,. \qquad \eee For  $C^+$ and $C^-$ of the
form (\ref{Csieg+}) and (\ref{Csieg-}), respectively, we have
\bee\nn && T_0(U,\,Y\,| \,X_0) =\,\,\,\\ \nn &&\ls\ls  \int d^M
\gx\,d^M\gl\,  c^+(\gx)  c^-(\gl) \exp i\Big (
X_0^{AB}(\gx_A\gx_B-\gl_A\gl_B) +   Y^A
(\gx_A-\gl_A)-U^C(\gl_C+\gx_C)
\Big). \eee So we obtain from (\ref{d4Y}) \bee\label{norma}\ls\ls
Q_0\sim\nn \int d^M\,\gx d^M\,\gl\,\,\,\,
c^+(\gx)\,\,c^-(\gl)\,\,\, \varepsilon\left( -\gl-\gx \, ,\,
\f{1}{2i}\left(\f{\p}{\p \gx} -\f{\p}{\p\gl}\right) \right) \
\gd^M(\gx-\gl)
\,\,\sim\rule{40pt}{0pt}\\
\int d^M\,\gx  \left( \varepsilon\left( -2\gx   ,
\f{i}{2}\f{\p}{\p \gn} \right) c^+(\gx+\gn)\,c^-(\gx-\gn)
\right)\Big|_{\gn=0}
\sim
\int d^M\,\gx  c^+(\gx)\, \varepsilon\left( -2\gx    ,
\f{i}{2}\f{\p}{\p \gx} \right) c^-(\gx )\, .\,\,\rule{10pt}{0pt}
\eee As expected, the result is independent of $X_0$. Up to a
rescaling of arguments it reproduces the alternative expression
for the charge obtained in \cite{cur}\,.

The important improvement of the form of the current (\ref{Q})
compared to (\ref{clo}) is that it allows us to introduce
singularities (fluxes) in the spinning variables
$W_\ga,\,\,W_\da,\,\, \gY^\ga\,,\gY^\da\,,$
$\Z^{\ga\gb},\Z^{\da\db} $ that can  bring in a proper singularity
into the charge (\ref{oldnew}) which, as we show in Section
\ref{conserv}, is needed  to reproduce HS currents of \cite{GSV}
in Minkowski space-time.

Finally, let us note that the generalized stress tensor can also
be constructed from arbitrary frequency fields as follows. Let
fields $\C_j (\gY|\Z)$ of frequencies  $\eta_j = \pm 1$ satisfy
the equations \bee \label{adholpm} \left ( \f{\p}{\p \Z^{ A B}} +
\,i\,\eta_j\, \f{\p^2}{\p \gY^ A \p \gY^B}\right ) \C_j(\gY|\Z)
=0\,. \eee Then the generalized stress tensor \bee\nn
\,\,T(\U,\gY|\Z)=\C_j\Big(\ga_j(\gY-\U))|\Z\Big)\,\,\C_k\Big(\gb_k(\U+\gY))|\Z\Big)\,\qquad
\Big(\,(\ga_j)^2=\eta_j,\,\, (\gb_k)^2=-\eta_k\,\Big) \, \eee
satisfies (\ref{unfol2uy}). Note however that the charge (\ref{Q})
built of regular parameters (\ref{par}) and fields of equal
frequencies, that are holomorphic in the same Fock-Siegel space,
vanishes because an integration surface can be deformed to the
respective infinity in the imaginary coordinates $\Z^{AB}$ where
the fields $\C_j (\gY|\Z)$ vanish. For the fields of opposite
frequencies such a deformation is not possible, that results in a
non-vanishing charge (may be, after an appropriate $\Hh-$shift).

\subsection{Appropriate classes of solutions }\label{class}
The   formula (\ref{norma}) for the charge  was obtained   under
assumption that the integrals  under consideration are
well-defined. The following four options are most significant
\begin{tabular}{ll}
(i)&$c^\pm(\xi)\in L_2(\mathbb{R}^M)$ ,\rule{0pt}{15pt}
\\
(ii)&$c^\pm(\xi)\in L_2^q(\mathbb{R}^M)$ , \rule{0pt}{20pt}
\\
(iii)&$c^\pm(\xi)\in S(\mathbb{R}^M)$ ,\rule{0pt}{20pt}
\\
(iv)&$c^\pm(\gx)\in S_{1/2}(\mathbb{R}^M)$ and
$c^\mp(\gx)\in S^\prime_{1/2}(\mathbb{R}^M)$ . \rule{0pt}{20pt}\\
&\rule{0pt}{10pt}
\end{tabular}

The cases (i)-(iii) are appropriate\footnote{Recall that
$L_p^q(\mathbb{R}^M)$ is the Sobolev space, and $S(\mathbb{R}^M)$
is the  Schwartz space of infinity differentiable functions
$f(\xi)$ that decrease at infinity  with all their derivatives
faster than any  multi-degree of $\f{1}{|\gx_j|}$.}, respectively,
for the cases of  $\gvep=1$ (electric charge), degree $q$
polynomial  $\gvep\left(\f{\p}{\p \gx} \right)$ and generic
polynomial $\gvep\left( -2\gx,\f{i}{2}\f{\p}{\p \gx} \right)$. The
case (iv)  is appropriate to pair a generalized function with a
test one. This is relevant to  the analysis of the composition
formulae for $\D$-functions in Subsection \ref{Composition}. As
mentioned  in Section \ref{Unfolded dyn}, in all these cases
$C^+(\gY|\Z)$ (\ref{Csieg+}) is analytic    and $C^-(\cY|\cZ)$
(\ref{Csieg-}) is anti-analytic in $\gY\,$ and $\, \Z$  provided
that $\Z\in \,\, \Ds_M^{(\Hh)}(\Z)$.

Then the generalized bilinear stress tensor (\ref{T_H}) is
\bee\label{TH}
T_\Hh{} ( U,\, Y|\Z)= C^+( Y\!-\! U|\Z)\,C^-{}( Y\!+\! U|\Z-i\Hh)=\rule{100pt}{0pt}\\
\nn\! \int\! d^M\!\gl d^M\gx c^+(\gx)c^-(\gl)\exp i \Big(\hbar
\gx_A\gx_B \Z^{AB}\!\! -\hbar \gl_A\gl_B
(\Z^{AB}\!\!-\!i\Hh^{AB}) +\!\gx_A(Y^A\!-\!U_A)    \!-\! \gl_B(
U^B\! \!+\!Y^B) \Big). \eee

Its Fourier transform (\ref{Fourier}) is 
\bee\label{FH}
f_\Hh{} ( W,\, Y|\Z)\sim  \rule{300pt}{0pt}\\
\nn \ls \int\! d^M\!\chi
c^+\Big(\f{-W+\chi}{2}\Big)c^-\Big(\f{-W-\chi}{2}\Big)\exp i
\Big(-\hbar W_A\chi_B \Z^{AB}\!\! +\hbar
\f{i}{4}(W+\chi)_A(W+\chi)_B  \Hh^{AB} +\! \chi_A\! Y^A  \!
\Big). \eee

Note, that for all cases (i)-(iv) \be
\label{cpmfh}c^+\Big(\f{-W+\chi}{2}\Big)c^-\Big(\f{-W-\chi}{2}\Big)\in
S^\prime_{1/2}(\mathbb{R}^{2M})\,. \ee As a result, since $\Z\in
\,\, \Ds_M^{(\Hh)}(\Z)$, $f_\Hh(W,\, Y\, |\Z)$ (\ref{FH}) is
integrable over $\mathbb{R}^M(W)$ $\forall (\Z,Y)\in \,\,
\Ds_M^{(\Hh)}(\Z)\times \mathbb{R}^M\,$. Moreover, the exponential
factor  with $\Z\in \,\, \Ds_M^{(\Hh)}(\Z)$ guarantees that
$f_\Hh(W,\, \gY\, |\Z)$ is analytic in $\gY$ and $\Z$.

As an illustrative example, let us consider functions of the form
\be\label{cpm}
c^\pm(\gx)=P^\pm(\gx)\,\exp(-E^\pm{}^{AB}\gx_A\gx_B) \ee with some
polynomial $P^\pm(\xi)$ and positive definite symmetric
$E^\pm{}^{AB}$, that allows explicit analysis. Evidently, the
integrals (\ref{norma}) converge absolutely for any polynomial
parameter $\gvep$. Evaluating the Gaussian integral, we have \bee
\label{pm} \ls f_\Hh(\W,\,\gY\, |\Z)&=& (2\pi)^{-M/2}\int
d^{\,M}U\,\exp\left(-i\,\,\,  \W_C\,U{}^C\right)\,\,\,\,\,\\ \nn
&& \ls\ls P^+\Big(i\f{\p}{\p U}\Big)\det{}^{-\half}(-i\Z+E^+)
\exp\Big(-
\f{1}{4}(-i\Z+E^+)^{-1}{}_{AB}(U^A-\gY^A)(U^B\!-\!\gY^B)\Big)\\
\nn &&\ls\ls\ls P^-\Big(i\f{\p}{\p
U}\Big)\det{}^{-\half}(\Hh\!+\!i\Z\!+\!E^-) \exp \Big(-
\f{1}{4}(\Hh\!+\!i\Z+E^-)^{-1}{}_{AB}(U^A\!+\!\gY^A)(U^B\!+\!\gY^B)\Big)\,.
\nn \eee The real parts $\Re(\Hh+i\Z+E^-)$ and $\Re(- i\Z+E^+)$
are positive definite provided that $\Z$ belongs to the Siegel
$\Hh-$~strip. Since from $\mathcal{F}\in \ZIGM$ follows
$\mathcal{F}^{-1}\in \ZIGM$ (see e.g. \cite{Mumford}),
$\Re\big((\Hh+i\Z+E^-)^{-1}\big)$ and
$\Re\big((-i\Z+E^+)^{-1}\big)$ are also positive definite. Hence
$f_\Hh(\W,\,\gY\, |\Z)$ (\ref{pm}) is holomorphic in its arguments
for $\Z\in   {{\Ds_M^{(\Hh)}}}$. Note that the "additional"
analyticity of $f_\Hh(\W,\,\gY\, |\Z)$  in $\W$ takes place for
any $c^\pm (\xi)\in S^{1/2}_{1/2} (\mathbb{R}^M)$, which is the
case for (\ref{cpm}).

\section{From  $\M_4$ to  Minkowski space}
\label{conserv}
\subsection{Idea of construction}
\renewcommand{\U}{ {U}}
\renewcommand{\W}{ {W}}

\label{idea}

In \cite{cur}, it was conjectured  that the charge (\ref{Q4})
should reproduce the HS  charges (\ref{Q3}) in Minkowski space
$M^4$ \cite{GSV} by an appropriate reduction to $M^4\subset \M_4$.
Since the charge in $\M_4$ contains four integrations versus three
in the Minkowski space, the naive reduction with the fourth
integration  over a cyclic spin variable in $\M_4$ gives zero
because the cycle is contractible. To make the cycle
noncontractible, a singularity (flux) should be introduced in the
spinning space.  As we demonstrate now, this can be done using the
generalized current (\ref{varpi}), which result was hard to
achieve starting from the original expression (\ref{clo}).

As explained in Section \ref{unfold}, the embedding of the $4d$
Minkowski space-time into $\M_4$ is conveniently described in the
language of  two-component complex spinors. In these terms,
$\W^A=(\W^\ga,\W^\da)$ and $\U^A=(\U^\ga,\U^\da)$. It should be
stressed that the complex structures of the Siegel space and of
two-component spinors are different. Correspondingly, since both
the real and imaginary parts of  the complex variables $\Z^{AB}$
are real symmetric matrices, in  terms of two-component complex
spinors we have \bee \nn \Z^{\za\zb} = (\Z^{\ga\gb} , \Z^{\ga\da}
,  \Z^{\da\db})= ( X^{\ga\gb} +i\X^{\ga\gb}, X^{\ga\da}
+i\X^{\ga\da}, {X}^{\da\db}+i\X^{\da\db}) \eee with \be\nn
\overline{X^{\ga\gb}}=X^{\da\db},\quad
\overline{X^{\ga\db}}=X^{\gb\da}\q
\overline{\X^{\ga\gb}}=\X^{\da\db},\quad
\overline{\X^{\ga\db}}=\X^{\gb\da}\,. \ee Note that
$\Re\Z^{\ga\gb}= \Re X^{\ga\gb}- \Im \X^{\ga\gb}$,
$\Im\Z^{\ga\gb}= \Im X^{\ga\gb}+\Re \X^{\ga\gb}$, {\it etc.}
Analogously we introduce $\gY^A=(\gY^\ga,\gY^\da).$

Let us choose the integration surface $\mathbf{ \Sigma}^{8}$ in
the form \bee \label{Sigdecomp3} \mathbf{ \Sigma}^{8}=
{\sigma}^3(\Z^{\ga\db})\times   
\gs^1(\Z^{\ga\gb} , \Z^{\da\db},\gY-\gY_0)  \times \gs^4( W)\,,
\eee where ${\sigma}^3(\Z^{\ga\db})$ is a three-dimensional
surface in the complexified Minkowski space ${}^{\mathbb{C}}M^4$,
$\gs^1(\Z^{\ga\gb} , \Z^{\da\db},\gY)$ is a one-dimensional cycle
in  the "spinning" subspace, $ \gs^4( W)\subseteq \mathbb{R}^4(W)$
is a four-dimensional surface   and $\gY_0$ is a free parameter.

An elementary calculation then shows that the pullback of the
differential form (\ref{varpi}) to the integration surface
$\mathbf{\Sigma^8}$ (\ref{Sigdecomp3}) gives
\bee\label{varpi3}\ls \varpi^{2M}(g)\big|_{\mathbf{
\Sigma}^{8}}\sim
  d\Z{}_{\ga\gga^\prime}\!\!\wedge d\Z{}^{\ga {\gb}^\prime}\!\!\wedge
d\Z{}^{\gb \gga^\prime} \wedge d\gsing(\W,\gY|\Z)\wedge d^4\W\,\,
g(\W,\gY |\Z) \W_\gb \W_{\gb^\prime}\,\Big|_{\mathbf{
\Sigma}^{8}}\, , \eee where \bee \label{zeta} \displaystyle
\gsing(\W,\,\gY\,|\Z)=\, \W_\gm \W_\gn \,\Z{}^{\gm\gn} -
{\W}_{\gm^\prime} {\W}_{\gn^\prime} \,\Z{}^{\gm^\prime \gn^\prime}
- \W_\gm \,\,\gY^{\gm} + {\W}_{\gm^\prime} \,\,\gY^{\gm^\prime}.
\eee

The key observation is that for any $\gY_0 $ the function
$\gsing(\W,\gY-\gY_0\,|\Z)$ solves (\ref{unfol2_Fur}) and is
independent of $\Z^{\ga \db}$. This allows us to use $\gsing$ to
introduce a singularity in a way independent of the complexified
Minkowski coordinates $\Z^{\ga\db}$.

To obtain Minkowski charge we set \be\label{zetah}
g_\gsing(\W,\,\gY\,|\Z)=
\gsing^{-1}(\W,\gY|\Z)\,g(\W,\,\gY\,|\Z)\,. \ee Since the
functions $\gsing(\W,\gY\,|\Z)$ and $g(\W,\,\gY\,|\Z)$ solve the
equations (\ref{unfol2_Fur}) , the same is true for
$g_\gsing(\W,\,\gY \,|\Z )$ (\ref{zetah}) away from singularities.
From (\ref{varpi3}) we have \bee\label{zuzuzuz}
\int\limits_{{\mathbf{\Sigma^8} }} \varpi^{2M}(g_\gsing) \sim
\int\limits_{\mathbf{\Sigma^8}} d\Z{}_{\ga\gga^\prime}\wedge
d\Z{}^{\ga {\gb}^\prime}\wedge d\Z{}^{\gb \gga^\prime}\,\wedge
\f{d\gsing  }{\gsing  }\wedge d^4\,\W\, \, g(\W,\,\gY\,|\Z)
\,\W_\gb \,\W_{\gb^\prime}\,. \eee The idea is to choose
one-dimensional cycle such that $\dis{\f{d\gsing  }{\gsing  }\,=i
\,d\phi}, $ where $\phi\in [0,2\pi)$ to be  a real coordinate on
$\sigma^1$. This can be done  as follows
\bee\label{syc1}\gs^1(\Z^{\ga\gb} ,
\Z^{\da\db},\gY)=\rule{200pt}{0pt}\\ \nn \rule{ 0pt}{20pt}\ls\{ \Z^{\ga\gb}= \gr S
{}^{\ga}{}^{\gb} \exp(i\phi) , \quad \Z^{\da\db}= \gr  S
{}^{\da}{}^{\db} \exp(i\phi) , \quad   \gY^\ga = \gr  S{}^\ga
\exp(i\phi),  \quad  \gY^\da =\gr  S{}^\da \exp(i\phi)\}, \eee
where $\gr>0$ is a real parameter and at least some of
parameters $S {}^ {\ga}{}^{\gb}$, $ S {}^{\da}{}^{\db}$,  $ S{}^
{\ga},$ and $S{}^{\db}$ are non-zero.

 For this choice  we
obtain \bee\label{gsinggs1} \gsing(\W,\,\gY\,|\Z)\big|_{ {\gs}^1}
 &=& \gr\exp(i\phi)P(W), \eee
 {where}
\bee\label{PW}  P(W)&=& W{}_\ga W{}_\gb  S {}^{\ga\gb} -  W{}_\da
W{}_\db S {}^{\da\db}-
   W{}_\ga  S{}^{\ga} +  W{}_\da S{}^{\da} \,.
\eee Hence\,\, $\dis{\f{d\gsing  }{\gsing  }\,=i \,d\phi}$\,,\,
and  a residue  is at $\dis{\gr=0}$ \,\,  for $P(W)\ne0$.

The subtlety that the integrand of the right side of
(\ref{zuzuzuz}) is not defined at   $P(W)=0$ does not affect the
result for any $g(\W,\,\gY\,|\Z) $  integrable over $\mathbb{R}^4
(W)$ because $P(W)$ cancels out in (\ref{zuzuzuz}) and the variety
$P: \,P(W)=0$ has measure zero. As a result, we obtain
\bee\label{otvetec} \ls Q  \sim \int\limits_{{\mathbf{\Sigma^8} }}
\varpi^{2M}(g_\gsing) = 2i\pi\ls \int\limits_{{\sigma}^3
\times\mathbb{R}^4 }\!\! d^3\,\Z^{\gb\db}\wedge\!
d^4\W\,\,\,g(\W,\,\gY_0\,|\Z^{\ga\da})\W_\gb \,\W_{\gb^\prime}\,,
\eee where $\dis{
    d^3\Z^{\gb\db} =d\Z_{\ga\da}\wedge d\Z^{\ga{\db}}\wedge d\Z^{\gb\da}\,.}$

For \bee\label{biling} g(\W,\,\gY\,|\Z)= \,\eta(\W,\,\gY\,|\Z) f
(\W,\,\gY\,|\Z)\,, \eee where $\eta(\W,\,\gY\,|\Z)$   is a
polynomial solution (\ref{par}) of (\ref{unfol2_Fur}) while $f
(\W,\,\gY\,|\Z)$ is the Fourier transform  (\ref{Fourier}) of the
generalized stress tensor $T (\U,\,\gY\,|\Z)$ we have \bee\nn\ls
Q\sim \int\limits_{{{ \sigma^3\times \mathbb{R}^4 } }}\!\!
 d^3\,\Z^{\gb\db}\wedge\! d^4\W
\, \, \eta(\W,\gY_0|\Z^{\ga\da}) f (\W,\gY_0|\Z )\W_\gb \W_{\gb^\prime}\,\,\sim\\
\label{otvetec2}
 \sim\!
\int\limits_{{\sigma^3 }}\! d^3\,\Z^{\gb\db} \f{\p^2}{\p U^\gb \p
U^\db} \,\left(\,\eta\,\big(i\f{\p}{\p\,U^C} , \,\gY_0\,| \Z
^{\ga\da} \big) T(U,\gY_0|\Z)\right)\big|_{U=0}\,, \eee which is
just the anticipated expression for conserved charge in Minkowski
space.

To give precise meaning to  this construction it remains to
identify ${\sigma}^3(\Z^{\ga\db})\subset{}^{\mathbb{C}}M^4$, to
choose appropriate Siegel strip $\Hh$, replacing the generalized
stress tensor $T (\U,\,\gY\,|\Z)$ by $T_\Hh (\U,\,\gY\,|\Z)$
(\ref{T_H}) bilinear in the massless fields $C^\pm$. This is done
in the next Subsection.

\subsection{Integration cycle}
\label{intc}

Let a positive definite matrix $\Hh\in \M_M$ be chosen in such a
way that $\Hh^{\ga\gb}=\Hh^{\da\db}=0$. Let
${\sigma}^3={\sigma}_\Hh^3(\Z^{\ga\db})$ be a real
three-dimensional surface in a real subspace of
${}^{\mathbb{C}}M^4$ defined by the condition $\Im \Z^{\ga\db}
=\gn \Hh^{\ga\db}$ , where $0<\gn<1$ is a free parameter. Note,
that with this choice, both $\Im\Z^{\ga\db}$ and $\Hh^{\ga\db}
-\Im\Z^{\ga\db}$ are positive definite for  $\forall\Z^{\ga\db}\in
\sigma_\Hh^3 $.

Let $ \sigma^1(\Z^{\ga\gb},\Z^{\da\db},\gY)$ be chosen in the form
(\ref{syc1}) such that both
\newcommand{\SYY}{\gY_0}
$$\mathcal{N}^{AB}=\Big(\Im\big( \gr S_0{}^
{\ga}{}^{\gb} \exp(i\phi)\big)\,,\, \gn\Hh^{\ga\db}\,,\,
\Im\big(\gr S_0{}^{\da}{}^{\db} \exp(i\phi)\big) \Big)$$ and
$$\Hh^{AB}-\mathcal{N}^{AB}$$
are positive definite, which is true for sufficiently small $\gr$
because this is the case for $\gr =0$.

For this choice we obtain that $\gs_\Hh^3\times\gs^1\subset
{\Ds_M^{(\Hh)}}\times \mathbb{C}^4(\gY)$.

Now let $T=T_{\Hh}(\U,\,\gY\,|\Z)$ be the stress tensor
(\ref{T_H}) built of the massless fields $C^\pm$ 
with the help of a positive definite regulator matrix $\Hh$. Let
$f_\Hh(\W,\,\gY\,|\Z)$ in (\ref{biling}) be   the Fourier
transform (\ref{Fourier}) of $T=T_{\Hh}(\U,\,\gY\,|\Z)$. If
$C^\pm$ are of the form (\ref{Csieg+}), (\ref{Csieg-}) with
functions $c^\pm$ from one of the classes (i)-(iv) of Subsection
\ref{class}, then  the pullback of the function $g(\W,\,\gY\,|\Z)$
(\ref{biling}) to the integration surface $ {\mathbf{
\Sigma}^{8}}$ (\ref{Sigdecomp3}) is integrable over $\mathbb{R}^4
(W)$ and the considerations of the previous subsection are true.
Therefore    (\ref{otvetec2}) acquires the form
\bee\label{otvetec1}\ls Q=Q_\Hh \sim \int\limits_{{\sigma_\Hh^3
}}\! d^3\,X^{\gb\db} \f{\p^2}{\p U^\gb \p U^\db}
\,\left(\,\eta\,\big(i\f{\p}{\p\,U^C} , \,\gY_0\,| \Z ^{\ga\da}
\big) T_\Hh(U,\gY_0|\Z)\right)\big|_{U=0}\,.\nn \eee
In the limit
$\Hh\to0$, this gives for   $\eta$ (\ref{par}) \bee \label{finY}
Q_0 \sim \int\limits_{{\sigma_0^3 }} d^3\,X^{\gb\db} \f{\p^2}{\p
U^\gb\p U^\db}\, \left( \gvep\Big(-i\f{\p}{\p\,U^C} ,
\,\gY_0+\,\,i\,X^{AB}\f{\p}{\p\,U^B}\Big) T_0(U,\,\gY_0
\,|X)\right)\Big|_{U=0}\, .\eee Recall that according to
(\ref{T_H}) \bee\nn T_0{} (U,\,\gY|X)=
C^+(\gY-U|X)\,C{}^-(\gY+U|X)\,. \eee Up to a rescaling of
variables,  for $\gY_0=0 $ this gives the conserved charge in
Minkowski space of \cite{GSV} for  polynomial  $\gvep(\W,\,\gY)$.

Let us stress that $Q $ (\ref{finY}) is $\SYY$ independent, \ie $
\f{\p}{\p \SYY^A}Q =0\,, $ because the variation over $\SYY^A$ is
equivalent to a local variation of the integration cycle away from
singularities. For $\gvep=const$ and $T$  of the form
(\ref{T_H}) this gives the following  identity
\bee\label{epsconst} \ls \ls 0=\! \int\limits_{{\sigma^3 }}\! d^3
X^{\gb\db} \f{\p^2}{\p U^\gb \p U^\db} \Big(\!
C^+(\gY\!-\!\U|X)\f{\p}{\p \gY^A}C^-(\gY\!+\!U|X) \!+ \!
C^-(\gY\!+\!U|X)   \f{\p}{\p \gY^A}C^+(\gY-\!U|X)
\Big)\,\Big|_{U=0}\,\,\,
\eee which will be used in the further analysis.

\section{Integral evolution formulae}
\label{Dfunc}

\subsection{$\D$-functions}
$\D-$functions  of the massless field equations (\ref{oscal}) and
(\ref{ofer}) in $\M_M$ were introduced in \cite{Mar} as their
singular solutions resulting from the integral representation
(\ref{cbfopmX+}) with $\dis{c^\pm = \mp i\pi^{-\f{M}{2}}}$ (in
this section we set $\hbar=1$), \be  \label{cD-} \D^+(\Z) =-
\f{i}{ \pi^M} \int\limits_{\mathbb{R}^M} d^M \gx\,\, \exp  i (
\gx_A\gx_B \Z^{AB}), \ee \bee\label{d} \D^-(\cZ)=-\D^+(-\cZ)
=\overline{\D^+ (\Z)} \, \eee and \bee\nn \D(\Z,\cZ) = \D^+ (\Z)
+\D^- (\cZ )=\D^+ (\Z)-\D^+ (-\cZ)\,. \eee

By construction, the functions $\D^- (\cZ)$, $\D^+ (\Z)$ and
$\D(\Z,\cZ)$ solve the equations of motion (\ref{oscal}). For $\Z$
in the upper  Siegel half-space $\ZIGM$, (\ref{cD-}) gives \be
\label{Dsol}
\D^+ (\Z) =
-{i}{\pi^{-\f{M}{2}}}{\dete}^{-1}\,, \ee where \be\label{dete}
\dete^2={\det (-i\Z)} \ee defines a multidimensional hyperelliptic
surface and $\dete$ is chosen to be holomorphic for $\Z=X+i\X \in
\ZIGM $ (\ie positive definite $\X^{AB}$) and to be positive real
for purely imaginary $\Z$, \ie $X=0$. As shown in \cite{Mar}
\be\nn
\D^+ (\Z)\Big |_{\X\to 0} =-\f{i}{\pi^{\f{M}{2}}} \exp\f{i\pi
I_X}{4} \f{1}{\sqrt{|\det (X)|}}\Big |_{\X\to 0}\,, \ee where
$I_X$ is the inertia index of the matrix $X^{\za\zb}$\,, \ie $ I_X
= n_+ - n_- \,, $ where $n_+$ and $n_-$ are, respectively, the
numbers of positive and negative eigenvalues of $X^{\za\zb}$.

{}From (\ref{d}) and  (\ref{Dsol}) it follows that \be\nn
\D^- (\cZ) =
{i}{\pi^{-\f{M}{2}}}{\cdete}^{\,-1}\,, \ee where $\cdete$ is
complex conjugated to $\dete$ (\ref{dete}).

The dependence of the $\D-$functions on $\gY$, $\cY$ is
reconstructed via the unfolded equations  (\ref{dgydgyh+}) and
(\ref{dgydgyh-}), respectively. In particular, \bee \label{D^+}
\D^+(\gY|\Z) =  \f{-i}{ \pi^M} \int d^M\xi\,  \exp   i (
\xi_{\za}\xi_{\zb} \Z^{\za\zb}\,+\gY^A\xi_A )
,  
\eee leading to \bee \label{dy} \D^+ (\gY|\Z) =  \f{-i}{
\pi^{M/2}}
\,\,s^{-1} \,\exp ( - \f{ i}{4} \Z_{AB}\gY^A \gY^B )\q
\Z_{AB}\Z^{BC} =\delta_{A}^{C}\,. \eee Notice that $\D^+ (\gY|\Z)$
(\ref{dy}) behaves as $-{2^M i} \delta^M(\gY)$ at $\Z\to 0$.

\subsection{Evolution formula in $\M_4$}
\newcommand{\chp}{{x^\prime{}}}
\newcommand{\chii}{{x {}}}
The obtained results allow us to give precise meaning to the
integral formula (\ref{CD}) in $\M_4$ by defining the integration
measure as corresponding to the  electric charge case in Minkowski
space. This guarantees that the restriction of so defined integral
representation to the Minkowski space correctly reproduces the
dynamics of massless fields.

Namely, we apply formula (\ref{finY}) with the generalized stress
tensor (\ref{T_H}) at $\Hh\to 0$, where $C^- (Y|X)$ is the
restriction of $C^- (\cY|\cZ)$ to the real subspace while
$C^+(\gY\,|\Z)$ is replaced by $\D^+ (\gY - \cY\,|\Z-\cZ)$ with
$\cY$ and $\cZ$ interpreted as parameters. (Recall that if $\Z_1,
\Z_2 \in \ZIGM$ then $\Z_1-\cZ_2 \in \ZIGM$ as well, including the
case, where either $\Z_1$  or $\Z_2$ is real, belonging to the
boundary of $\ZIGM$.)
\renewcommand{\SYY}{Y_0}
Let the resulting conserved charge (\ref{finY}) with
$\gvep=\half$, which is a function of $\cY$ and $\cZ$, be denoted
as $\widetilde{C}^-(\cY\,| \cZ)$. We obtain \bee \label{finYDC}
\ls \widetilde{C}^-(\cY\,| \cZ) = \half\int\limits_{{\sigma^3 }}
d^3 \XX^{\gb\db} \f{\p^2}{\p  U^\gb \p  U^\db}
\Big({\D}^+(\SYY - U-\cY|\XX-\cZ)C^- (\SYY+ U|\XX) \Big) \Big|_{
U=0},\,\, \eee where \,$\SYY$ is a free parameter,\,\,
  $\gs^3\subset M_4( X)$ \, is\, an
Euclidean  three-dimensional  subspace $\mathbb{R}^3$ of Minkowski
space, \be\label{sigma3}\gs^3(X)=\{\T_{\ga\db} X^{\ga\db}=t_0\},
\ee where a positive-definite matrix $\T_{\ga\db}$ describes the
time arrow, the time evolution parameter is $t=\half \T_{\ga\db}
X^{\ga\db}$ and $t_0$ is its value associated with the chosen
space surface. For example, for $\T_{\ga\db}=\delta_{\ga\db}$, $
X^{\ga\db}=t\gd^{\ga\db}+ x{}^i \sigma_i^{a\db}$, where
$\sigma_i^{a\db}$ are Pauli matrices.

The key observation is that from (\ref{finYDC}) it follows  by
virtue of (\ref{epsconst}) that all derivatives of
$\widetilde{C}^- (\cY\,| \cZ)$ with respect to $\cY$ are  related
to those of $C^-{}   (\gY|\XX)$ with respect to $ \gY$  just in
the same way as $\widetilde{C}^- (\cY\,| \cZ)$ and $C^-{}
(\gY|\XX)$ in
(\ref{finYDC}), \ie 
\bee \label{AAA} \f{\p}{\p \cY{}^{A_1}}\dots \f{\p}{\p
\cY{}^{A_m}}
 \widetilde{C}^- (\cY\,| \cZ)
= \rule{50pt}{0pt}\\ \nn \half\int\limits_{{\sigma^3 }} d^3
\XX^{\gb\db} \f{\p^2}{\p  U^\gb \p  U^\db}
\Big({\D}^+(\SYY - U -\cY|\XX-\cZ)\,  \f{\p}{\p  U^{A_1}}\dots
\f{\p}{\p  U^{A_m}} C^-{}   (\SYY+ U|\XX) \Big) \Big|_{ U=0}\,.
\eee

Let us now show that   from (\ref{finYDC}) it   follows that \bee
\label{evol} \widetilde{C}^-( 0\,| \,\cZ)\Big|_{\bar{\Z}\in \gs^3}
&=&C^-( 0\,| \,\cZ)\Big|_{\bar{\Z}\in \gs^3}\,. \eee Using  that $
d^3 X^{\gga\dga}\Big |_{\gs^3} =d^3 x\, \T^{\gga\dga}\Big
|_{\gs^3}$\,\,,\, we obtain from (\ref{finYDC}) \bee
\label{evolpr} \ls\widetilde{C}^- (0|\cZ) =
i\half\int\limits_{\sigma^{3}} d^3\chp\, \Big( C^-{}  (0|\XX)\,
\dot{\D}{}^+(0|\XX\!-\cZ) - \dot{C}{}^-{}  (0|\XX)\,
{\D}{}^+(0|\XX\!- \cZ)\Big) ,\,\,  \eee where we have taken into
account that the functions  $\D^+ $ and $ C^- $ satisfy the
unfolded equations (\ref{dgydgyh+})  and (\ref{dgydgyh-}),
respectively, together with the facts that
$$
\frac{\partial}{\partial  U^{A}} {\D}^+(- U  | X-\cZ )\Big|_{
U=0}=0
$$ as a consequence of (\ref{dy}) and
\bee\label{dot} \T^{\ga\da}\f{\p^2}{\p U {}^\ga \p   U  {}^\da }
C^\pm(  U  |X ) = \pm i \dot{C}{}^\pm(  U  |X )\,, \qquad
\dot{C}{}^\pm(U  |X ) = \f{\p}{\p t }C^\pm ( U  |X )\,. \eee

To prove (\ref{evol}), it is convenient to use the complex
notation for the space coordinates $\chii^{j}$
\cite{Mar} \be\nn \chii=\chii^1+i\chii^2\,,\qquad \bar{\chii} =
\chii^1-i\chii^2 \,, \ee such that $
   d\chii_1 \wedge d \chii_2\,\wedge d\chii_3=\f{1}{2i}
 d\chii_3 \wedge d\chii \wedge d\bar{\chii}\,$.

The combinations of  $\xi_\ga$ dual to the coordinates
$\chii^3,\quad \chii ,\quad \bar{\chii}$ \cite{Mar} \be \label{k3}
k_3 =\xi_1\bar{\xi}_{\dot{1}}-\xi_2\bar{\xi}_{\dot{2}} \,,\qquad
\overline{k} =2 \xi_1 \bar{\xi}_{\dot{2}} \,,\qquad k=2
\bar{\xi}_{\dot{1}} \xi_{2} \ee map $\mathbb{R}^4 /{Z}_2$ on
$\mathbb{R}^3$, i.e. $k_3,k$ and $\overline{k}$ can take arbitrary
values. The leftover ambiguity in the integration variables
$\xi_\ga$ for a given $k_i$ is the overall phase factor $\xi_\ga
\to \exp{\half i\varphi}\, \xi_\ga$, $ \varphi \in [0,2\pi )$.
(Recall that $\xi_\ga$ is identified with $-\xi_\ga$.) We set \be
\label{cyc} \exp{i\phi} = 2 \f{\xi_1 \xi_2}{k}\,,\qquad
\exp{-i\phi} = 2 \f{\overline{\xi}_{\dot{1}}
\overline{\xi}_{\dot{2}}}{\overline{k}}\,. \ee The integration
measures are related as follows \cite{Mar} \be\label{mes4} dk_3
\wedge dk \wedge d\bar{k} \wedge d \phi =-8 i
(\xi_1\bar{\xi}_{\dot{1}}+\xi_2\bar{\xi}_{\dot{2}}) d\xi_1\wedge
d\bar{\xi}_{\dot{1}}\wedge d\xi_2\wedge d\bar{\xi}_{\dot{2}}\,.
\ee The map (\ref{k3}), (\ref{cyc}) from $\mathbb{R}^4 /{Z}_2$
associated with the  variables $\xi_\ga$ to $\mathbb{R}^3\times
S^1$ described by the variables $k_i$, $\phi$ is non-degenerate
except for the expected singularity at $\xi_\ga =0$. Note that
\be\label{mar4} \xi_1\bar{\xi}_{\dot{1}}+\xi_2\bar{\xi}_{\dot{2}}
= \sqrt{k \bar{k}+k_3^2} = \sqrt{k_1^2 +k_2^2+k_3^2 }\,. \ee

For any $C^+$ (\ref{Csieg+}) and  $C^-$ (\ref{Csieg-}) we have
using  formulae (\ref{k3})--(\ref{mar4}) \bee \label{D-M0}
{C}{}^\pm (0|t ,\, x)=
\f{1}{\pi^2}\int d^4 \gx\,\, c^\pm(\gx) \exp \pm i(t  k_0+ x^3 k_3
+ x \bar{k} +{\bar{ x} } k)\,,\eee \bee \label{D-M}
\dot{C}{}^\pm(0|t ,\, x)= \f{1}{\pi^2}\int d^4\gx\, \,(\pm
\,i\,k_0)\, c^\pm(\gx) \exp \pm i(t  k_0+\, x^3 k_3 + x \bar{k}
+{\bar{ x} } k)\,.\eee Using  (\ref{D-M0}) and (\ref{D-M}), we
obtain from (\ref{evolpr}) \bee \nn
\ls\ls\ls\widetilde{C}^-{}(0\,|t_0,\chii) =
 \f{i}{4\pi^6}\,
\int\limits_{\sigma^{3}}   d\chp_3 \wedge d\chp \wedge
d\bar{\chp}\, \int\!d\gk_3 \wedge d\gk \wedge d\bar{\gk} \wedge d
\psi \,dk_3 \wedge dk \wedge d\bar{k} \wedge d \phi
\rule{20pt}{0pt} \\ \nn \ls\ls\f{i}{64 }\left(\!\f{1}{
\gk_0}\!+\!\f{1}{k_0}\!\right) c^-(\gk,\psi) \exp -i(
t_0\gk_0+\chp^3 \gk_3 \!+\!\chp\bar{\gk}\!+\!{\bar{\chp}}\gk )
\exp i((\chp^3\!-\!\chii^3 ) k_3 +(\chp\!-\!\chii
)\bar{k}+{(\bar{\chp}\!-\!\bar{\chii} )}k ) =\\ \nn
\f{1}{\pi^2}\int d\gk_3 \wedge d\gk \wedge d\bar{\gk} \wedge d
\psi \f{i}{ 8 \gk_0}\,\,c^-(\gk,\psi)\, \, \, \exp
i(-t_0\gk_0-\chii^3 \gk_3 -\chii \bar{\gk} -\bar{\chii} \gk )\,
\,=\, \, C^-(0|t_0,\chii)\,.\rule{35pt}{0pt} \label{evolprDCM2}
\eee Analogously,  from (\ref{AAA}) it   follows that \bee
\label{evoldt} \widetilde{C}^-{}_{A_1\dots A_m}(\, 0\,| \,\cZ)
\Big|_{\bar{\Z}\in \gs^3} &=&  C^-{}_{A_1\dots A_m}(\, 0\,| \,\cZ)
\Big|_{\bar{\Z}\in \gs^3}\,, \eee where  $\dis{{C}^- (\cY\,|
\cZ){}_{A_1\dots A_m}\equiv \f{\p}{\p \cY{}^{A_1}} \dots \f{\p}{\p
\cY{}^{A_m}} {C}^-( \cY\,| \,\cZ)} $. {}This is equivalent to
\bee\nn \widetilde{C}^-{}( \cY\,| \,\cZ) \Big|_{\bar{\Z}\in \gs^3}
=  C^-{}( \cY\,| \,\cZ) \Big|_{\bar{\Z}\in \gs^3}\,. \eee Because
the dependence on $\cY$ fully reconstructs the dependence on $\cZ$
by virtue of the unfolded equation (\ref{dgydgyh-}) from here it
follows that \bee\label{full} \widetilde{C}^-(\cY\,| \cZ)=
{C}^-(\cY\,| \cZ)\,, \eee that proves the evolution formula \bee
\label{evolformula}
 {C}^-(\cY\,| \cZ)
= \half\int\limits_{{\sigma^3 }} d^3 \XX^{\gb\db} \f{\p^2}{\p
U^\gb \p  U^\db}
\Big({\D}^+(\SYY - U-\cY|\XX-\cZ)C^- (\SYY+ U|\XX) \Big) \Big|_{
U=0} .\rule{10pt}{0pt} \eee

Analogously one can see that \bee \label{evolformula+}
{C}^+(\gY\,| \Z)= \half\int\limits_{{\sigma^3 }} d^3 \XX^{\gb\db}
\f{\p^2}{\p  U^\gb \p  U^\db}
\Big({\D}^-( {\SYY} + U-\gY|\XX-\Z)\, 
C^+{}  ( {\SYY}- U|\XX) \Big) \Big|_{ U=0}.\rule{10pt}{0pt} \eee

Note, that from here the usual formulae for spin 0 and spin 1/2
fields  follow \bee \nn \ls {C}^- (0|\cZ) =
i\half\int\limits_{\sigma^{3}} d^3\chp\, \Big( C^-{}  (0|\XX)\,
\dot{\D}{}^+(0|\XX\!-\cZ) - \dot{C}{}^-{}  (0|\XX)\,
{\D}{}^+(0|\XX\!- \cZ)\Big) \,\,,  \eee \bee \nn \ls {C}^-_\ga
(\,0|\cZ) = \int\limits_{\sigma^{3}} d^3\chp\, \, \T^{\gb
\db}\,C^-_\gb  (\,0|\XX)\, {\D}{}^+_{\ga\db}(\,0|\XX\!-\cZ) \,\,,
\eee where ${\D}{}^+_{\ga\db}(\,\gY| \Z)=\f{\p}{\p
\gY^{\ga}}\f{\p}{\p  \gY^{\db}}{\D}{}^+ (\,\gY| \Z)$, etc.

\subsection{Evolution formula in the Fock space}
\label{EVOLFOR} Now let us show that the same result can be
obtained  for any $M$ from (\ref{d4Y}) at $\gvep=\f{i}{(2i)^M}$ by
the integration  over the $\gY$ variables as \be\label{DCY}
{C}^-(\cY|\cZ) = \f{i}{(2i)^M}\int\limits_{\gY^\prime=Y^\prime}
d^M \gY^\prime\,\, \D^+(\gY^\prime - \cY|\Z^\prime-\cZ)
C^-(\gY^\prime |\Z^\prime)\, \ee with any $\Z^\prime$. Indeed,
substituting \be\nn C^-(\gY  |\Z )=\int d^M \xi \,\,c^-(\xi) \exp
-i\big(\xi_A \gY^{  A} +\xi_A \xi_B \Z^{  A  B}\big) \ee into
(\ref{DCY}) and using the expression (\ref{D^+}) for $\D^+$, we
obtain
\bee\nn
&&\f{i}{(2i)^M}\int d^M Y^\prime\,\,\, \D^+(Y^\prime -
\cY|\Z^\prime-\cZ) \,C^-(Y^\prime |\Z^\prime)\, =\\ \nn \ls\!\!
\f{1}{(2\pi i)^M} &&\ls\ls\int d^M Y^\prime d^M \xi d^M \lambda
\,\, c^-(\xi) \exp -i\big(\lambda_A (\cY-Y^{\prime})^A +\lambda_A
\lambda_B (\cZ -\Z^{\prime})^{ A B} +\xi_A Y^{\prime A} +\xi_A
\xi_B \Z^{\prime A  B}\big)=
\\ &&\nn
\int  d^M \gx\,\, c^-(\gx) \exp -i\big(\gx_A  \cY{}^A   +\gx_A \gx_B
\cZ{}^{AB}\big)\,=\,{C}^-(\cY  |\cZ ). \eee

\subsection{Composition properties of the $\D$-functions}
\label{Composition} Let us consider
\be\label{compprop}
C^+(\Z|\gY)=\D^+(\Z|\gY)\mbox{ \quad and \quad}
C^-(\cZ|\cY)=\D^-(\cZ-\Z^\prime|\cY-\gY^\prime),
\ee where
$\Z^\prime\in\ZIGM$ and $\gY^\prime\in \mathbb{C}^M$ are free
parameters. One can see, that
\bee\nn \ls C^-(\cZ|\cY) =
 \int d^M\xi\, c^-(\gx)
\exp   -i ( \xi_{\za}\xi_{\zb} \cZ{}^{\za\zb} \!+\!\cY{}^A\xi_A ),
\quad c^-(\gx)=\f{i}{ \pi^M}
 \exp   i ( \xi_{\za}\xi_{\zb} \Z^\prime{}^{\za\zb}\!+\!\gY^\prime{}^A\xi_A )
\eee and
\bee\nn C^+(\Z|\gY)= \int d^M\xi\, c^+(\gx) \exp   i (
\xi_{\za}\xi_{\zb} \Z^{\za\zb}+\gY^A\xi_A )\,,\quad
c^+(\gx)=\f{-i}{ \pi^M}\,.
\eee Since $c^-(\gx)\in S_{1/2 }$ and
$c^+(\gx)\in S^\prime{}_{1/2}$ for any $\Z^\prime\in \ZIGM$, we
can pair   $C^+$ and $C^-$ (\ref{compprop}) in the evolution
formula (\ref{evolformula+}) to obtain
\bee \label{D+-}
 {\D}^-(\cY\,| \cZ)
= \half\int\limits_{{\sigma^3 }} d^3 \XX^{\gb\db} \f{\p^2}{\p
U^\gb \p  U^\db}
\Big({\D}^+(- U-\cY|\XX-\cZ)\D^- (  U|\XX) \Big) \Big|_{ U=0}.
\rule{10pt}{0pt} \eee Analogously from  (\ref{evolformula}) it
follows that \bee \label{D-+} {\D}^+(\gY\,| \Z)=
\half\int\limits_{\sigma^3 } d^3 \XX^{\gb\db} \f{\p^2}{\p  U^\gb
\p  U^\db} \Big({\D}^-(   U-\gY|\XX-\Z)\, \D^+{}  ( - U|\XX) \Big)
\Big|_{ U=0}.\rule{10pt}{0pt} \eee These formulae express the
composition properties of the $\D$-functions.

\subsection{Solutions associated with the $\D$-functions}

Finally let us note that one can use the formula (\ref{dy}) to
generate new solutions of the HS equations as follows. Since
$\D^+(\gY|\Z)$ solves the HS field equations for any $\gY$,
\bee\nn \f{1}{\sqrt{\det (i\Z)}} \,\exp \Big( - \f{ i}{4} \Z_{AB}
(\gY^A+uV^A)(\gY^B+uV^B) \Big) \,, \eee where $V^A$ is any
constant vector and $u$ is a constant parameter, also does.
Integrating this expression at $\gY=0$ with respect to $u$ with
some weight $\rho(u)$, we arrive at the following  set of
solutions of the equation (\ref{oscal}) \be \label{fdetsolCOM}
C(0|\Z) = f\Big(V^A V^B\Z_{AB}\Big)\det{}^{-\half}(i\Z) \,, \ee
where  $f$ is an arbitrary double differentiable function. One can
check directly that $C(0|X)$ (\ref{fdetsolCOM}) solves
(\ref{oscal}).

\bigskip
\section{Higher rank conserved currents}
\label{Multinear} In \cite{tens2}, we have introduced higher rank
fields in $\M_M$. Positive- and negative-frequency  rank $r$
fields
  satisfy the unfolded equations of the form
\be \label{hfe} \left\{ \, \f{\p}{\p X^{AB}}  \pm\, i\,
\eta_{k\,l}\, \f{\p^2}{\p   Y{}_{\,k}{}^{B}\p   Y{}_{\,l}{}^{A}}
\right\} C^\pm(Y\,|X)=0\,, \ee where $k,l=1\ldots r$ and
$\eta_{kl}$ is a positive definite symmetric form. In \cite{tens2}
it was also argued that rank $r$ fields in $\M_M$ can be
interpreted as resulting from the reduction of rank 1 fields in
$\M_{rM}$ to $\M_M$ diagonally embedded into $\M_{rM}$ via \bee
\label{emb} X^{(A,k)(B,l)} = \eta^{kl} X^{AB}\,. \eee A particular
realization of higher rank fields is provided by products of lower
rank fields. For example, a product of two rank 1  fields \be
\nn 
C(Y_1,Y_2\,|X)=C(Y_1|X)C(Y_2|X) \ee gives a rank 2  field.

Analogously, we introduce a higher rank generalization of the
current equation (\ref{unfol2_Fur})
\be \label{unfol2r_Fur}
\left\{ \, \f{\p}{\p X^{AB}}  + \,\, W_k{}_{(A} \f{\p}{\p
Y{}_{\,k}{}^{B)}} \right\} g(W,\,Y\,|X)=0\,,
\ee where
$g(W,\,Y\,|X)$ takes values in $\M_M \times \mathbb{R}^{r\,M}
\times  \mathbb{R}^{r\,M}$ with  coordinates $\,  X^{AB},\,
W_j{}_A,Y_{j\,}{}^A\,$, $A,\,B=1,...M$, $j=1,...r$. Again,
particular solutions of the higher rank current equation are
provided by the products of lower rank currents. The higher rank
current equation allows us to derive  multilinear conserved
currents. Indeed one can easily see that the $2rM$ differential
form
\bee\label{varpi2r} \varpi_{2rM}\Big(g(W,\,Y\,|X)\Big)\,=\,
\left(d\,W_k{}_A \wedge \Big(\,W_k{}_B{} d\,X^{AB} -\,d\,Y_k{}^A
\Big)\right)^{r\,M}\,\,g(W,\,Y\,|X)
\eee is closed provided that
$g(W,\,Y\,|X)$ satisfies  (\ref{unfol2r_Fur}). As a result, on
solutions of (\ref{unfol2r_Fur})  the charge
\be \label{Qr} Q^r=
\int_{\Sigma^{2rM}} \varpi_{2rM}(g )
\ee is independent of local
variations of a $2rM$-dimensional surface $\Sigma^{2rM} $. In
particular, it is time-independent\,\, hence providing a conserved
charge.

The formula (\ref{varpi2r}) gives rise to conserved currents for
$g(W,\,Y\,|X) $ of the form (\ref{eta_f}) where the ``symmetry
parameter" $\eta(W,\,Y\,|X)$ is of the form \be \label{parr}
\eta(W,\,Y\,|X) = \varepsilon(W_j{}_A,\,Y_k{}^C
\,-\,X^{CB}\,W_k{}_B )\, \ee and $f(W,\,Y\,|X)$ is a   solution of
(\ref{unfol2r_Fur}) related to a multilinear``stress tensor"
$T(U,\,Y\,|X)$ via the Fourier transform \bee \label{FourierHr}
f(W,\,Y\,|X)= (2\pi)^{-rM/2}\int d^{r\,M}U\,\,\,
\exp\left(-i\,\,\, W_j{}_C\,U^j{}^C\right)\, T(U,\,Y\,|X)\,, \eee
\bee \label{fourierHr} T(U,\,Y\,|X)= (2\pi)^{-rM/2} \int
d^{r\,M}\,W\,\,\, \exp\left(i\,\,\, W_j{}_C\,U^j{}^C\right)\,
f(W,\,\,Y\,|X)\,. \eee

The equations (\ref{unfol2r_Fur}) are equivalent to the following
equations for $T$ \be \label{unfol2ruy}\left\{ \,\f{\p}{\p X^{AB}}
- i  \, \f{\p}{\p Y_j{}^{(A} } \f{\p}{\p U^j{}^{B)}} \right\}
T(U,\,Y|X) =0, \ee \ie $f(W,\,Y\,|X)$ satisfies
(\ref{unfol2r_Fur}) provided that $T(U,\,Y\,| X)$ satisfies
(\ref{unfol2ruy}) and vice versa.

A $2r-$multilinear ``stress tensor" $T(U,\,Y\,|X)$ can be
constructed analogously to the bilinear one \cite{cur} as follows
\bee \label{stress2r}
T(U,\,Y\,|X)=\prod_{j=1}^r\,C^+{}_j(Y_j-U_j|X)\,C^-{}_j(U_j+Y_j|X),
\eee where $C^\pm{}_1(Y|X),\,\dots,\,C^\pm{}_r(Y|X)$ are solutions
of positive- or negative-frequency rank 1   equations.

The higher rank form (\ref{varpi2r}) can be interpreted as the
pullback of the standard form (\ref{varpi}) in $\M_{rM}$ to the
diagonal subspace $\M_M\subset \M_{rM}$ (\ref{emb}). The
multilinear stress tensor (\ref{stress2r}) is nothing but the
bilinear stress tensor (\ref{stress2})  on the rank one solutions
in $\M_{rM}$, that result from rank $r$ solutions in $\M_M$
provided by the $r$-linear products of the rank 1  solutions in
$\M_M$.

The formula (\ref{stress2r}) gives rise to multilinear currents
built of free massless fields. Naively, the higher conserved
currents constructed from this stress tensor amount to algebraic
functions of the bilinear currents. This is indeed true if the
integration measure factorizes into a product of lower-rank
measures, that concerns both the integration surface and the
parameter function $\eta(W,Y|X)$ (\ref{parr}), but may not be true
e.g. for nonpolynomial $\eta(W,Y|X)$  that contains a singular
dependence analogous to (\ref{zetah}) in the case of Minkowski
bilinear current. Note that the resulting nontrivial currents
should be nonlocal from the perspective of usual Minkowski
space-time because the charge (\ref{Qr}) contains $2rM$
integrations instead of $2M$ in the rank 1  case. Specific
examples of higher-rank currents will be considered elsewhere.

Note that,   analogously to the consideration of  Sections
\ref{Unfolded dyn} and \ref{Bilinear}\,, all higher rank
constructions can be complexified. Namely, one can consider
holomorphic continuation $C(\gY_k^A|\Z ^{BC})$ of the fields
$C(Y_k^A|X^{BC})$, that take values in $\mathfrak{H}{}_M(\Z)\times
\mathbb{C}^{rM}(\gY)$.

Finally let us note that the higher rank system (\ref{hfe}) is
invariant under the group $O(r)\times \mathbb{R}$ that acts as
follows \be\nn
\gY^A_i\to \widetilde{\gY}^A_i = \exp({\phi})\, T_i{}^j \gY^A_j\q
\Z^{AB}\to \widetilde{\Z}^{AB}=\exp({2\phi})\, \Z^{AB}\,, \ee
where $\phi\in \mathbb{R}$ and $T_i{}^j \in O(r)$ leaves invariant
the metric tensor $\eta_{jk}$. This symmetry can be used for the
derivation of identities between solutions of the equations
(\ref{hfe}) using the fact that if two functions $C_1(\gY|\Z)$ and
$C_2(\gY|\Z)$ satisfy  (\ref{hfe}) and coincide at some $\Z=\Z_0$,
$C_1(\gY|\Z_0)=C_2(\gY|\Z_0)$, then they coincide for any $\Z$.

{}For example, the following identities hold for the
$\D^\pm$-functions \be\label{Riemann} \D^\pm(\gY_1|\Z)\ldots \D^\pm
(\gY_r |\Z) = \exp{(rM\phi)}\,
\D^\pm(\widetilde{\gY}_1|\widetilde{\Z}) \ldots \D^\pm
(\widetilde{\gY}_r |\widetilde{\Z})\,. \ee In particular, for the
rank 2  case we obtain
\be\label{Riemann2}
{2^M}\D^\pm(2\U|2\Z)\D^\pm(2\gY|2\Z) =
\D^\pm(\U+\gY|\Z)\D^\pm(\U-\gY|\Z)\,. \ee Identities
(\ref{Riemann})  are continuous analogues of the well-known
generalized Riemann identities of theta functions.

\bigskip

\section{Riemann theta functions as solutions of higher-spin equations}
\label{thetafunc}

\subsection{Theta functions  in the Fock-Siegel space }

A somewhat surprising property of the   massless field equations
formulated in the Fock-Siegel space  $\mathfrak{H}_M\times
\mathbb{C}^M$ is that
Riemann theta functions form their natural solutions. Indeed,  a
general positive-frequency solution (\ref{Csieg+}) periodic under
$\gY^A\to \gY^A +n^A$, $n^A\in \mathbb{Z}^M$ has the form \be
\label{genper} C^+(\gY|\Z)= \sum_{n^A\in \mathbb{Z}^M} c^+_n
\exp{i(\hbar\Z^{AB}(2\pi n_A )(2\pi n_B) +2\pi n_C \gY^C)}\,. \ee
With $c^+_n=1$ and $\hbar=\f{1}{4}\pi^{-1}$ this formula gives the
standard expression for the Riemann theta function \cite{Mumford}
\be \label{theta} \theta(\gY,\,\Z)= \sum_{n^A\in \mathbb{Z}^M}
\exp{i\pi(\Z^{AB}n_A n_B +2 n_A \gY^A)}\,. \ee Here the
complexified space-time coordinates $\Z^{AB}$ identify with the
complex period matrix that defines quasi-periods of
$\theta(\gY,\,\Z)$,
\bee\nn 
\theta(\gY+m\Z,\,\Z)= \exp{(-i\pi\Z^{AB}m_A m_B -2i\pi m_A
\gY^A)}\,\theta(\gY,\,\Z)\q
 m_A\in \mathbb{Z}^M\,.
\eee Also let us note that theta function is $\Z$ periodic in the
sense \be\nn \theta(\gY\,,\Z+\Z_{int})=\theta(\gY\,,\Z)\,, \ee
where $\Z^{AB}_{int}$ is any real symmetric  matrix with integer
elements and even diagonal elements.

The fundamental reason why theta functions solve the HS field
equations is that both HS theory \cite{F,BHS} and the theory of
theta functions \cite{Mumford} are based on the $Sp(2M)$ symmetry
and its Weyl-Heisenberg   extension which,  on the HS gauge theory
side, is just the HS symmetry. For example, as mentioned in
Introduction, for the case of $M=2$ the $\mathfrak{sp}(4)\sim
\mathfrak{o}(3,2)$ identifies with the conformal symmetry of
massless scalar and spinor in three space-time dimensions. In the
case of $M=4$, the appearance of the $\mathfrak{sp}(8)$ symmetry,
that acts on the infinite sets of all bosonic and all fermionic
massless fields, is a less trivial fact observed originally in
\cite{F}. The equation (\ref{dydy}), which is the simplest
$\mathfrak{sp}(8)$ invariant unfolded equation, was shown in
\cite{BHS} to describe properly massless fields of all spins in
four dimensions (a closely related argument was also given in
\cite{BLS}). On the other hand, it is well-known that theta
functions form a $\Gamma_{1,2} $-module, where
$\Gamma_{1,2}
$ is the Igusa subgroup of $Sp(2M|\mathbb{Z})$ \cite{Mumford}.

Thus, the fact that theta functions satisfy the same equations as
HS fields is not too mysterious once $Sp(2M)$ appeared in the HS
theory. Moreover, from the HS theory perspective, the
$\Gamma_{1,2} $ symmetry in the theory of theta functions is the
leftover symmetry of the continuous HS symmetry that leaves
invariant a particular solution of the HS field equations provided
by the theta function up to a phase. This class of solutions may
indeed play a distinguished role in the HS theory because
conserved currents constructed from such solutions, turn out to be
invariant under $\Gamma_{1,2}$.

The roles of the variables $\Z^{AB}$ and $\gY^A$ in the HS theory
and the theory of theta  functions is to some extent opposite. In
the HS theory, $\Z^{AB}$ are space-time variables while the
twistor variables $\gY^A$ play an auxiliary role at least in the
conventional field-theoretical picture. The indices $A=1,2\ldots
M$ for $M=2^k$ are interpreted as spinorial on the HS theory side.
In the theory of 
theta functions, $M$ identifies with genus $g$, the period matrix
$\Z^{AB}$ (usually denoted $\Omega^{AB}$ \cite{Mumford}) plays a
role of a parameter, while the dependence on $\gY^A$ (usually
denoted $z^A$) is of most interest. Note however, that in the
nonlinear HS theory it was realized since  nineties (see
\cite{Gol} and references therein) that the fundamental HS
dynamics is encoded in terms of the twistor variables $\gY^A$,
while the role of $\Z^{AB}$ is to visualize the HS dynamics in
terms of local events \cite{Mar}.

Many of the well-known properties of theta functions acquire a
nice interpretation in terms  of the HS symmetry  (\ref{gtr}) of
the fundamental unfolded equation (\ref{dgydgyh+}) which is a
multidimensional analogue of the Schrodinger equation. For
example, theta functions with characteristics $\,b^A \in
\mathbb{R}^M$, $a_A \in \mathbb{R}^M$ \bee\nn
\theta\big[{}^{\,a}_{\,b}\big](\gY,\,\Z)= \exp{\big(i\pi\Z^{AB}a_A
a_B +2i\pi a_A \gY^A+ {2i\pi a_A b^A}\big)}\,\,\,\,\theta(\gY+\Z
a+b,\,\Z)\\ \nn =\sum_{n^A\in \mathbb{Z}^M}
\exp{\big(i\pi\Z^{AB}(n_A+a_A)( n_B+a_B) +2i\pi (n_A+a_A)
(\gY^A+b^A)\big)}\, \eee that also solve (\ref{dgydgyh+}), result
from the action of the HS symmetry
(\ref{gtr}) 
with $\jj_A=2i\pi a_A$, $\jh^B=b^B$ and $\mu=\f{i}{4\pi}$ on the
theta function.

Consider the set of theta functions with equal characteristics $
a^A=b_A = 0$ or $\half$, $ \forall A=1,2\ldots M, $ \bee
\label{clth} \theta\big[{}^{\,a}_{\,a}\big](\gY,\,\Z)
=\sum_{n^A\in \mathbb{Z}^M} \exp{i\pi(\Z^{AB}(n_A+a_A)( n_B+a_B)
+2 (n_A+a_A) (\gY^A+a^A))}\,, \eee which consists of $2^M$
independent functions. It is easy to see that \bee\nn
\theta\big[{}^{\,a}_{\,a}\big](-\gY,\,\Z) = (-1)^{\sum_{A=1}^{M}
2a_A}\,\,\theta\big[{}^{\,a}_{\,a}\big](\gY,\,\Z)\,, \eee \ie
$2^{M-1}$ functions with an odd number of $a_A=\half $ are odd in
$\gY^A$ and $2^{M-1}$ functions with an even number of $a_A=\half
$ are even in $\gY^A$. In accordance with the normal relationship
between spin and statistics, the odd functions describe
half-integer spin massless fields while the even functions
describe integer spin massless fields (in the former case the
solution has to be multiplied by a Grassmann odd element). In
particular, the theta function (\ref{theta}) is a member of this
set with $a_A=0$, \ie it is bosonic.

The class of solutions $C^+(\gY|\Z)$ (\ref{genper}) of the
unfolded HS field equations in $\M_M$ is special in the sense that
$C(Y|0)$ is not a regular function of $Y$ as is usually assumed in
the unfolded HS analysis, but becomes a distribution {at real
$\Z$}.

Indeed, one can see that \be\nn \theta\big[{}^{\,a}_{\,a}\big]
(\gY ,\,0) = \sum_{n^A\in \mathbb{Z}^M}(-1)^{\sum_{A=1}^M 2a_A
n^A}\delta^M(\gY^A + a^A-n^A)\,. \ee

This formula means in particular, that
$\theta\big[{}^{\,a}_{\,a}\big](\gY,\,\Z)$ develops a singularity
at $\Z\to 0$. As such, it is analogous to the $\D$-function of the
massless field equations. In fact, it is the $\D$-function of HS
field equations for solutions with appropriate (anti)periodicity
conditions in $\gY$. In the limit in which the period of $\gY$
variables tends to infinity, \ie the Fourier series in
(\ref{theta}) is replaced by the Fourier integral,
$\theta(\gY,\,\Z)$  becomes the $\D$-function as it is obvious
from (\ref{D^+}). The counterpart of the evaluation formula
(\ref{DCY}) for periodic in $\Re \gY$ function $C^-(\cY|\cZ)$,
conjugated to $C^+(\gY|\Z)$ (\ref{genper}) is \be\label{TCY}
{C}^-(\cY|\cZ) = 
\int\limits_{[01)^M} d^M Y^\prime\,\, \theta( Y^\prime -
\cY|\Z^\prime-\cZ) C^-( Y^\prime |\Z^\prime)\, \ee with  any
$\Z^\prime$. (The direct proof is analogous to that of Subsection
\ref{EVOLFOR}.) The
 evolution formula for $C^+(\gY,\Z)$   is obtained with the help of
 $\theta^-(\cY,\cZ)=\overline{\theta(\gY,\Z)}$.

Also, let us note that the generalized Riemann identities for
theta functions \cite{Mumford} are discrete analogues of the
identity (\ref{Riemann}) for $\D^+$.
For example, the generalized Riemann identity 
\be\nn \sum_{2a_A\in (\mathbb{Z} / 2\mathbb{Z} )^M}
\theta\big[{}^{ \,a}_{\, 0}\big](2\gY,\,2\Z)
\theta\big[{}^{\,a}_{\,0}\big](2\U,\,2\Z)
=\theta{(\gY+\U\,,\Z)}\theta{(\U-\gY\,,\Z)} \ee can be derived
along the same lines as (\ref{Riemann2}).

\subsection{Theta functions and massless fields in Minkowski space}

In the case of $M=2$, eq.~(\ref{clth}) gives solutions for
massless scalar and spinor in three dimensions. In the case $M=4$,
it describes the superpositions of massless fields of all spins in
four dimensions. Let us stress that direct identification of theta
functions with solutions of $4d$ massless field equations turns
out to be so simple just because, in the $Sp(8)$ invariant
framework, massless fields of all spins turn out to be involved.
This is a manifestation of the general feature that linear and
nonlinear field equations for massless fields of all spins are in
a certain sense simpler than those for specific lower spins.

The reduction of the theta function solutions to a definite spin
in the Minkowski subspace of $\M_M$ is more subtle. To this end,
let us first of all make precise the relationship between Majorana
indices $A,B,\ldots$ and two-component indices $\ga,\gb\ldots$ and
$\da,\db\ldots$.  Let $A^A= (A^1, A^2, A^3, A^4)$. We set for
two-component vectors $\tilde{A}$ \be\nn \tilde{A}^\ga = (A^1 +i
A^3, A^2 +i A^4)\q \tilde{A}^\da = (A^1 -i A^3, A^2 -i A^4)\,. \ee
The inverse relations are \be\nn A^1 = \half (\tilde{A}^1
+\tilde{A}^{1^\prime} )\,,\quad A^2 = \half (\tilde{A}^2
+\tilde{A}^{2^\prime} )\,,\quad A^3 = \f{1}{2i} (\tilde{A}^1 -
\tilde{A}^{1^\prime} )\,,\quad A^3 = \f{1}{2i} (\tilde{A}^2 -
\tilde{A}^{2^\prime} ). \ee Introducing \be\nn \nu_1 = n_1 -i
n_3\,,\quad \nu_2 = n_2 -i n_4\,,\quad {\nu}_{1^\prime} = n_1 +i
n_3\,,\quad {\nu}_{2^\prime} = n_2 +i n_4\,, \ee which describe
points  with integer coordinates on the complex plane, we observe
that \be\nn n_A A^A = \nu_\ga \tilde{A}^\ga +\nu_\da
\tilde{A}^\da\,. \ee Also we introduce the complex characteristics
\be\nn \ga_1 = a_1 -i a_3\,,\quad \ga_2 = a_2 -i a_4\,,\quad
{\ga}_{1^\prime} = a_1 +i a_3\,,\quad {\ga}_{2^\prime} = a_2 +i
a_4\,. \ee

As a result, the theta function with characteristics (\ref{clth})
can be rewritten as \be\nn
\theta\big[{}^{\,a}_{\,a}\big](\gY,\,\Z) =\ls
\sum_{\Re\nu^A,\Im\nu^A\in \mathbb{Z}^4} \ls
\exp{i\pi(\Z^{\ga\gb}\mu_\ga \mu_\gb +2 \Z^{\ga\db}\mu_\ga
{\mu}_\db+ \Z^{\da\db}{\mu}_\da {\mu}_\db + 2\mu_\ga
\tilde{\gY}^\ga+2{\mu}_\da \tilde{\gY}^\da)}\,, \ee where \be\nn
\mu_\ga = \nu_\ga +\ga_\ga\q  {\mu}_\da =  {\nu}_\da +{
\ga}_\da\,. \ee The solutions of spin $s$ field equations in the
complexified Minkowski space, that result from
$\theta\big[{}^{\,a}_{\,a}\big](\gY,\,\Z)$, are \be\nn
C_{\ga_1\ldots\ga_{2s}}({\Z}) =(2i\pi)^{2s}
\sum_{\Re\nu^A,\Im\nu^A \in\, \mathbb{Z}^4} \mu_{\ga_1}\ldots
\mu_{\ga_{2s}} \exp{\big(2i\pi \Z^{\ga\db}\mu_\ga
{\mu}_\db\big)}\,, \ee \be\nn C_{\da_1\ldots\da_{2s}}({\Z}) =
(2i\pi)^{2s}\sum_{\Re\nu^A,\Im\nu^A \in\, \mathbb{Z}^4}
{\mu}_{\da_1}\ldots  {\mu}_{\da_{2s}} \exp{\big(2i\pi
\Z^{\ga\db}\mu_\ga  {\mu}_\db\big)}\,. \ee
\section{Conclusion}
\label{conc}

The main technical result of this paper is  the definition of the
proper integration measure for the conserved charges of $4d$
massless fields in terms of an integral in the ten-dimensional
matrix space $\M_4$. This allowed us not only to reproduce the
known HS charges in Minkowski space starting from the
ten-dimensional matrix space $\M_4$ but also to give the integral
evolution formulae for massless fields via the $\D$-functions in
$\M_4$. The precise integration prescription is given in terms of
the Siegel  upper half-space $\ZIGM$ \cite{Siegel} of complex
$M\times M$ symmetric matrices $\Z^{AB}=\Z^{BA}$ with positive
definite imaginary parts. More generally, we observe that massless
fields are most conveniently described in terms of the Siegel
space with complex matrix coordinates. In this setup, positive--
and negative--frequency solutions are described, respectively, as
holomorphic and antiholomorphic functions in the
 Siegel  upper half-space $\ZIGM$.
The systematic reformulation of the HS fields in the Siegel space
leads to a number of surprising conclusions.

One is that the unfolded form of the classical massless field
equations studied in this paper distinguishes between positive and
negative frequencies, \ie particles and antiparticles, the
property that is usually delegated to the quantization
prescription. We interpret this intriguing observation as the
important indication that the unfolding procedure is able to
describe quantization. It is worth to note that the unfolded
equations themselves have a form of a multidimensional Schrodinger
equation.

Another important observation is that Riemann theta functions
provide a natural class of periodic solutions of the $4d$ massless
field equations. This fact is a consequence of the $Sp(8)$ and HS
symmetries of the massless field equations. The setup of unfolded
HS field equations is convenient for the analysis of properties of
theta functions. For example, Riemann-type identities
\cite{Mumford} can be derived by analysing solutions of the rank
$r$ equations (\ref{hfe}) analogously to the analysis of massless
field $\D$-functions in Section \ref{Dfunc}.

Theta functions provide non-zero solutions of massless field
equations that are invariant  up to a phase factor $\sqrt[8]{1}$
under the  transformations from the {Igusa} group
$\Gamma_{1,2}\subset Sp(2M,\mathbb{Z})$ \cite{Mumford}. As a
result, theta functions may indeed play a distinguished role in
the HS theory as highly symmetric nontrivial solutions because the
observables constructed from such solutions, like, e.g., conserved
currents, turn out to be invariant under $\Gamma_{1,2}$.

An intriguing possibility would be if such a solution can be
identified with a nontrivial vacuum of the HS theory. Although any
such a solution breaks down the Lorentz invariance to some its
discrete subgroup, such a breakdown can be compatible with the
observations if the scales of the periods of the vacuum solutions
are large enough. In that case, the breakdown of the Lorentz
invariance may have cosmological implications.

The fact of natural appearance of theta functions in the HS gauge
theory is expected to shed light on a still mysterious
relationship of HS gauge theories with String Theory and
integrable systems.

\bigskip

\section*{Acknowledgement}
We are grateful to Mikhail Soloviev for very helpful comments and
Augusto Sagnotti for warm hospitality at Scuola Normale Superiore,
Pisa in October-November  2007 where an important part of this
work was done. This research was supported in part by INTAS Grants
03-51-6346 and No 05-7928, RFBR Grant No 05-02-17654, LSS No
4401.2006.2. One of us (M.V.) acknowledges a partial support from
the Alexander von Humboldt Foundation Grant PHYS0167.

\newcounter{appendix}
\setcounter{appendix}{1}
\renewcommand{\theequation}{\Alph{appendix}.\arabic{equation}}
\addtocounter{section}{1} \addcontentsline{toc}{section}{
\,\,\,\,\,\,Appendix. Ring of solutions of  massless equations}

\medskip
\section*{Appendix. Ring of solutions of  massless equations}
\label{Ring} As observed in Section \ref{Bilinear}, the space of
solutions of the first-order unfolded current equation
(\ref{unfol2_Fur})  forms a commutative associative algebra.  A
less trivial fact is that the space of solutions of the
second-order unfolded equation (\ref{unfol2uy}) can also be
endowed with the structure of associative commutative algebra as
follows.

Let us introduce  the following commutative and associative
product $\tudasuda$ on the space of fields  $A(\vy|X)$:
\be\label{trans} \left(A{\tudasuda} B\right)(Y|X)= A(\vy|X)
\exp\left\{ -2\mu \frac{\overleftarrow{\partial}}{\partial
Y^\vga}X^{\vga\vgb} \frac{\overrightarrow{\partial }}{\partial
Y^\vgb} \right\} B(\vy|X). \ee The space of solutions of the
unfolded system (\ref{dydy}) is closed under the product
$\tudasuda $\,,\, \ie given two solutions $A,\,\,\,B$ of
(\ref{dydy}), $A{\tudasuda} B$ is  its new solution as is easy to
see by straightforward substitution of (\ref{trans}) into
(\ref{dydy}).

The meaning of the product $\tudasuda$ is quite simple. It
corresponds to the usual product of the ``initial data" $A(Y|0)$
and $B(Y|0)$ of the respective problems as follows from
(\ref{trans}) at $X^{AB}=0$. So, it is not too surprising that the
space of solutions forms such an Abelian algebra. It is
remarkable, however, that the product $\tudasuda$ has the simple
and constructive form (\ref{trans}) in $\M_M$. The integral
version of the formula (\ref{trans}) with $\mu=1$ reads as
\be\nn 
\left(A{\tudasuda} B\right)(Y|X)=\frac{det |X^{AB}|}{(2\pi)^{M}}
\int dW\, dV A(Y+W|X) B(Y+V|X) \exp {-\half W^A V^B X_{AB}}\,, \ee
where the matrix $X_{AB}$ is  inverse to $X^{AB}$.

The generalization to  rank $r$ fields in $\M_M$ \cite{tens2} is
straightforward
\bee\nn 
{\tudasuda}_{\otimes^r} = \exp\left\{-2 \sum_{i=1}^r
\frac{\overleftarrow{\partial}}{\partial Y_j^A}X^{AB}
\frac{\overrightarrow{\partial }}{\partial Y_j^B}\right\} . \eee

\end{document}